\documentclass[prc,showpacs]{revtex4}
\usepackage{epsfig,graphicx,float,color}
\usepackage{amssymb}

\newcommand{\be}{\begin{equation}} 
\newcommand{\ee}{\end{equation}}
\newcommand{\bc}{\begin{center}}
\newcommand{\ec}{\end{center}}

\begin{document}

\title{Phase diagram for a Cubic Consistent-Q Interacting Boson
  Model Hamiltonian: signs of triaxiality}
\author{L. Fortunato$^{1,2,3}$,  C.E. Alonso$^4$, J.M. Arias$^4$,
  J.E. Garc\'{\i}a-Ramos$^5$ and A. Vitturi$^{2,3}$} 
\affiliation{$^1$ ECT*, Strada delle Tabarelle 286, I-38050 Villazzano (TN), Italy \\
$^2$ Dipartimento di Fisica ``G.Galilei'', via Marzolo 8, 
I-35131 Padova, Italy \\
$^3$ INFN, Sezione di Padova, via Marzolo 8, 
I-35131 Padova, Italy \\
$^4$ Departamento de F\'{\i}sica At\'omica, Molecular y Nuclear,
Facultad de F\'{\i}sica, Universidad de Sevilla, Apartado~1065,  
41080 Sevilla, Spain \\ 
$^5$ Departamento de F\'{\i}sica Aplicada, Universidad de Huelva,
21071 Huelva, Spain} 

\begin{abstract}
An extension of the Consistent-Q formalism for the Interacting
Boson Model that includes the cubic $(\hat{Q}\times \hat{Q}\times 
\hat{Q})^{(0)}$ term is proposed. The potential energy surface for the
cubic quadrupole 
interaction is explicitly calculated within the coherent state
formalism using the complete ($\chi-$dependent) 
expression for the quadrupole operator. The Q-cubic term is found
to depend  on the asymmetry deformation parameter
$\gamma$ as a linear combination of $\cos{(3\gamma)}$ and
$\cos^2{(3\gamma)}$ terms, 
thereby allowing for triaxiality.  The phase diagram of the model
in the large N limit is explored, it is  described the order
of the phase transition surfaces that define the phase
diagram, and moreover, the possible nuclear equilibrium shapes are established.
It is found that, contrary to expectations, there is only
a very tiny region of triaxiality in the model, and that the transition
from prolate to oblate shapes is so fast that, in most cases, the
onset of triaxiality might go unnoticed.
\end{abstract}

\maketitle

\section{Introduction}
\label{sec-intro}
The quadrupole operator plays a central role in Nuclear Physics
because it is essential in the description of nuclear deformation,
in the calculation of energy terms and in the evaluation of electromagnetic quadrupole
transitions and moments \cite{Bohr75}. Its presence is also of key importance
within the Interacting Boson Model (IBM) 
\cite{ibm,GK}, where its components are defined, in terms of $s$ and $d$ bosons, as 
\be
\hat{Q}_\mu=[d^\dag \tilde s+s^\dag \tilde d]^{(2)}_\mu +\chi [d^\dag \tilde
  d]^{(2)}_\mu .
\label{qop}
\ee
Various relevant operators (including scalar terms in the
Hamiltonian or the $E2$ electromagnetic transition operator) 
are usually built from $\hat{Q}$ by appropriate tensor couplings. Within the IBM a
useful and, at the same time, extremely simple Hamiltonian is the one of the 
Consistent-Q formalism (CQF) introduced some time  ago
by D.D. Warner and  R.F. Casten \cite{Warn83}. This model Hamiltonian
is formed by a one body monopole  
term, proportional to the number of $d$ bosons, plus a
quadrupole-quadrupole interaction among bosons.   
The CQF Hamiltonian can generate spherical,
deformed axially symmetric as well as deformed $\gamma-$unstable potential energy
surfaces \cite{ibm}. It was noted \cite{ibm} that for a general s-d IBM Hamiltonian
including up to two body terms, triaxial shapes are
forbidden (please, note that triaxiallity can be obtained with up to two-body
terms if g-bosons are included \cite{sdgPiet}). Within the standard
s-d IBM several authors have shown (for example in
Refs.~\cite{Hey,ThJE,Garc00a,Garc00b,ibm}, in ch. 6 of Ref. \cite{BB}, ch. 2.2 of
Ref. \cite{Bon} and in Ref.~\cite{Jolo04}) that the triaxiality can
be successfully introduced by adding three-body terms of
the type $[d^\dag d^\dag d^\dag]^{(L)}\cdot[\tilde d\tilde d\tilde
d]^{(L)}$. These terms, together with a CQF Hamiltonian, generate a
relatively broad region of triaxiality in the parameter space of the Hamiltonian.
Although this is a valid
way of generating triaxiality, it is not completely satisfactory because there is a
priori no reason to invoke such terms and moreover one moves away of
the desirable simplicity of the CQF framework. An alternative is to use
higher order terms in $\hat{Q}$ as, for instance,  the cubic
$(\hat{Q}\times \hat{Q}\times \hat{Q})^{(0)}$ interaction.
The explicit expression for the cubic order interaction reads, 
\be (\hat{Q}\times \hat{Q}\times \hat{Q})^{(0)}=\sum_{\nu,\mu}
\langle 2\nu 2-\nu | 00\rangle \langle 2\mu 2(\nu-\mu)|2\nu\rangle
\hat{Q}_\mu \hat{Q}_{\nu-\mu}\hat{Q}_{-\nu}, 
\ee 
where $\langle . . . .| . .\rangle$ stands for Clebsh-Gordan
coefficients. 
Here three quadrupole operators (with rank-2 tensorial properties) 
are coupled to give a scalar
term. This operator couples states with $\Delta \tau=3$ as well as
$\Delta \tau=1$, where $\tau$ is the $O(5)$ seniority quantum number.
  
Several studies have been carried out for disentangling the properties
of Hamiltonians containing $\hat{Q}\times \hat{Q}\times \hat{Q}$ terms.
On the one hand, P.~Van Isacker \cite{piet} has
studied the tensorial properties of the 
$\hat{Q}\times \hat{Q}\times \hat{Q}$
operator, within the IBM, and has shown that, even with $\chi=0$, this cubic
operator can give spectra and band structures that are qualitatively
similar to those usually associated with a SU(3) type of
symmetry (axial rotor).  
On the other hand, Rowe and Thiamova \cite{RoTi} have recently investigated the
spectrum generated by using the following Hamiltonian
\be
\hat{H}=\hat{\Lambda} +k(\hat{Q}\times \hat{Q}\times \hat{Q})^{(0)} \;,
\label{h3}
\ee
where $\Lambda$ is the Casimir operator of $O(5)$, reaching similar
conclusions to those of P.~Van Isacker: the spectrum obtained by increasing
the strength of the cubic term with $\chi=0$ displays the
properties of an axially symmetric rotor.  In their paper
and, more recently in Ref.~\cite{Thi}, it is shown that the
contraction of the O(6) algebra indicates that $\cos{(3\gamma)}$ is
an image of the cubic term and therefore the only possible stable
shapes are axially deformed. Following this reasoning, a quadratic term in the
$\hat{Q}$ cubic scalar, {\it i.e.} $(\hat{Q}\times \hat{Q}\times \hat{Q})^{(0)} \cdot
(\hat{Q}\times \hat{Q}\times \hat{Q})^{(0)}$, is necessary to generate potential energy terms
displaying a $\cos^2{(3\gamma)}$ behavior. While this is most
certainly true with their assumption on the form of the quadrupole
operator ($\hat{Q}_\mu=[d^\dag s+s^\dag \tilde d]^{(2)}_\mu$,
{\it i.e.} $\chi=0$), we find by explicit calculation that a term quadratic
in the cosine is already present in the matrix elements of the
$(\hat{Q}\times \hat{Q}\times \hat{Q})^{(0)}$ operator alone, if one includes the $\chi$
dependent term in the definition of the quadrupole operator as in
Eq.~(\ref{qop}). Therefore the possible onset of triaxiality can be studied at
the level of the cubic CQF Hamiltonian (to be defined below) without the need to
resort to  higher order terms, in agreement with the
already mentioned studies on the $[d^\dag d^\dag
d^\dag]^{(L)}\cdot[\tilde d\tilde d\tilde d]^{(L)}$ operators.

After this introductory section we introduce the cubic CQF Hamiltonian (Section
\ref{sec-c-cqf}) and then obtain the potential
energy surfaces (PES) generated by the $(\hat{Q}\times \hat{Q}\times \hat{Q})^{(0)}$
operator. In Section \ref{sec-ph-dia} we explore the phase
diagram and discuss the order of the phase transition
surfaces. Interesting limiting situations of this Hamiltonian are explored in 
Section \ref{sec-appli} and, finally, Section \ref{sec-conclu} is
devoted to summarize our conclusions.

\section{Cubic Consistent-Q Hamiltonian}
\label{sec-c-cqf}
The Consistent-Q formalism (CQF) is based on a simple
IBM Hamiltonian  \cite{Warn83} that allows to investigate not only the
three dynamical 
symmetries of the IBM-1, but also the transitional regions in between. 
We propose in this section an extension of the CQF by adding to the
original Hamiltonian a cubic combination of $\hat{Q}$ operators coupled to
zero angular momentum, 
\be
H=\xi \hat{n}_d -(1-\xi) \Bigl[ \frac{(\hat{Q}\cdot \hat{Q})}{N} +
k_3\frac{(\hat{Q}\times \hat{Q}\times \hat{Q})^{(0)}}{N^2} \Bigr] \;. 
\label{h23}
\ee
We call Eq. (\ref{h23}) the Cubic Consistent-Q Hamiltonian (CCQH) because we
consistently keep the same values of  
$\chi$ in all the quadrupole operators appearing in the above equation,
up to the cubic degree.  
As usual, the various terms have been divided by the appropriate power
of the boson number ($N$) in order to preserve the same kind of $N$
dependence in the large $N$ limit and also to make sure that, in this limit, each term 
does not go to a infinity value.
Although many other three body terms can be used (for instance the seventeen linear
independent three body terms discussed in Refs. \cite{Garc00a,Garc00b}
that have been successfully used in the interpretation of double
phonon anharmonicities using the IBM),  
this looks like the simplest one and it is easy to justify on
physical grounds as the first higher order interaction term in an 
expansion based on the quadrupole operator. The $\hat{Q}$ cubic 
term can be interpreted as a correction to the quadrupole-quadrupole
scalar product rather than an additional term.   

This Hamiltonian has a rich structure that will be analyzed in depth
in the following sections. Clearly, when $\xi=1$, one falls back into the
U$(5)$ spherical limit. When $\xi=0$ and $k_3=0$ one recovers the
deformed $\gamma-$unstable ($\chi=0$) and the axially deformed
($\chi=\pm \sqrt{7}/2$) IBM limits. For values $k_3 \neq 0$ there will
be a competition between different possible deformed shapes that could
produce, in principle, stable triaxiality. In order to analyze this
competition, and consequently the appearance of triaxiality, let us
first investigate the geometry produced by the Q-cubic term in the
Hamiltonian.

\subsection{Geometry of the $(\hat{Q}\times \hat{Q}\times \hat{Q})^{(0)}$ operator}
\label{subsec-qqq-geo}
One fundamental step in the investigation of the cubic Q term is the
connection of the $(\hat{Q}\times \hat{Q}\times \hat{Q})^{(0)}$ operator with geometry,
that can be obtained within the well-known intrinsic state
formalism. A coherent, or intrinsic, state \cite{ibm,GK} is defined as a properly
normalized application of the $N-$th power of a linear
combination of scalar and quadrupole boson creation operators to the vacuum,
namely,  
\begin{eqnarray}
\mid \beta,\gamma, N\rangle &=& \frac{1}{\sqrt{N!(1+\beta^2)^N}}\nonumber\\
&\times& 
\left[s^\dag + \beta \cos{\gamma}~ d_0^\dag + \frac{\beta}{\sqrt{2}}
\sin{\gamma}~(d_{2}^\dag +d_{-2}^\dag )\right]^N \mid 0 \rangle \;, 
\end{eqnarray}
where $\beta$ and $\gamma$ are related to the deformation and asymmetry
parameters, respectively. The main difficulty, when one deals with 
matrix elements of complicated operators within the coherent state formalism,  
is the length of the calculations. We have, therefore, set up a symbolic
computer code that can evaluate in an analytic fashion complicated
terms, keeping into account the non-commutative nature of the basic
operators \cite{Fort-unp}. This code adopts the technique introduced
in Ref. \cite{VaCh} of transforming the creation (annihilation)
operators into derivatives acting on the left (right). After having
carefully tested the procedure with several known quantities
\cite{BB}, we have obtained the following result for the matrix element   
\be
\langle \beta,\gamma, N \mid (\hat{Q}\times \hat{Q}\times \hat{Q})^{(0)}\mid
\beta,\gamma, N \rangle =\sum_i t_i \;,
\label{qqqform}
\ee
that is split for simplicity in one-, two-, and three-body terms
($i=1,2,3$) given by
\begin{widetext}
\begin{eqnarray}
t_1 &=&\frac{N}{14\sqrt{5}}~\frac{1}{(1 + \beta^2)} ~\left( 14 \chi (5 + 2\beta^2) - 
3 \chi^3 \beta^2 \right), \\
t_2 &=& \frac{N(N-1)}{49 \sqrt{5}}~\frac{3 \beta^2}{(1 +
  \beta^2)^2}~\left(14 \chi (14 + \beta^2) - 
   3 \chi^3 \beta^2 - \sqrt{14} (14
     + 11 \chi^2) \beta \cos{3\gamma} \right),\\
t_3 &=& \frac{N(N-1)(N-2)}{49 \sqrt{5}}~ \frac{4 \beta^3}{(1 +
  \beta^2)^3} \left( 42 \chi \beta - 
   \sqrt{14} (14 + 3 \chi^2 \beta^2) \cos{3\gamma} + 
\chi^3  \beta^3 (2 \cos^2{3\gamma}-1) \right).
\end{eqnarray}
\end{widetext}
To obtain the full matrix element these three terms have to be summed up, keeping in
mind that, in the intrinsic state formalism, only the leading order in
$N$ is correctly evaluated and, hence, only $t_3$ is meaningful \cite{Dusu05a,Dusu05b}. 

From these expressions it is clear that the dependence on $\gamma$
when $\chi=0$ is of the type $\cos{3\gamma}$ and therefore the cubic
operator with $\chi=0$ cannot generate triaxiality. 
The full expression, valid for any value of $\chi$, contains one term
proportional to $\cos{3\gamma}$ together with another proportional
to  $\cos^2{3\gamma}$. Therefore these contributions, in addition to a
CQF energy surface, might generate a triaxial minimum in the
potential energy surface. 

\begin{figure}[!t]
\begin{center}
\begin{tabular}{cc}
\includegraphics[width=9cm,clip=]{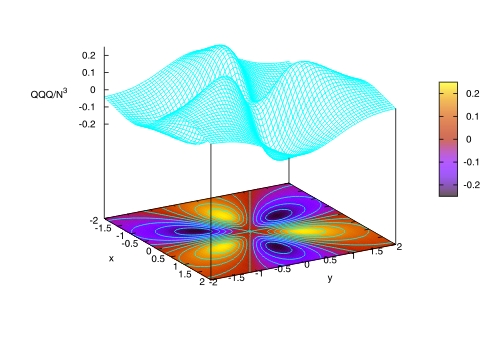} &
\vspace{-1cm}\includegraphics[width=0.32\textwidth, bb= 0 0 315 315]{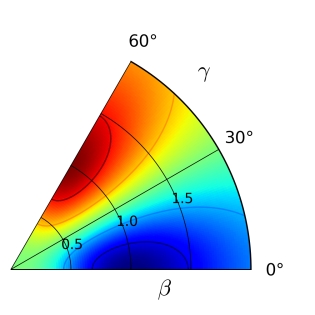} \\
\end{tabular}
\end{center}
\caption{(Color online) Left: surface (grid) and contours (colored) representing $(\hat{Q}\times
  \hat{Q}\times \hat{Q})^{(0)}/N^3$ in the large $N$ limit with $\chi=0$ as a
  function of $\{x,y\}$. The coordinates on the plane are $x=\beta
  \cos \gamma$ and $y=\beta \sin \gamma$. Right: the same function in the fundamental
 $0^\circ \le \gamma \le 60^\circ$  wedge with contours (slightly different coloring scheme).} 
\label{qqq}
\end{figure}

In order to analyze the energy surface (\ref{qqqform}),
we will  examine in Fig. \ref{qqq} the potential energy surface given
by  $(\hat{Q}\times \hat{Q}\times \hat{Q})^{(0)}/N^3$ in the large $N$ limit for $\chi=0$.
The upper left-part of the figure displays the value of the matrix element
on the vertical coordinate, while the lower left-part is its
projection on the plane with coordinates  
$x=\beta \cos \gamma$ and $y=\beta \sin \gamma$. The surface has
prolate minima at $\beta=1$ and $\gamma= \frac{2\pi}{3}n$ and oblate
maxima at $\beta=1$ and $\gamma= \frac{\pi}{3}(1+2n)$ with $n=0,1,2$.  
The contour map on the $x-y$ plane gives a color-coded projection of
the surface values, with black/blue representing the (prolate) minimum
and yellow/red representing the (oblate) maximum (see color bar on
the right).  For completeness sake we also give in the
right-part of the figure a contour plot  
in the more standard $(\beta, \gamma)$ polar coordinates, limited to
the  $0^\circ \le \gamma \le 60^\circ$ wedge with a slightly different coloring scheme. 
Of course the character of the minima/maxima can be interchanged by an overall sign
change. In conclusion, the cubic-Q term with $\chi=0$ only produces either prolate
or oblate minima. Triaxiality or $\gamma-$unstability  are not
allowed with the $\hat{Q}\times \hat{Q}\times \hat{Q}$ term if $\chi=0$.

\section{The phase diagram in the large N limit}
\label{sec-ph-dia}
The study of the shape of a quantum system, in particular of atomic
nuclei, proceeds through the study of the energy surface in the large $N$
limit. The shape is strictly defined in the thermodynamical limit. The
procedure starts by calculating the potential energy surface in the
large $N$ limit, then a minimization should be done, getting the
equilibrium value of the deformation parameters.

\subsection{Potential Energy Surface}
\label{sec-energy}
The energy per boson in the large N limit can be readily calculated from Eq.~(\ref{h23}):
\begin{widetext}
\begin{eqnarray}
E(\xi,k_3,\chi,\beta,\gamma)&=& \xi \frac{\beta^2}{1+\beta^2}- (1-\xi)\Biggl [
\frac{2}{7}~\frac{\beta ^2}{(1+\beta ^2)^2}\Big(\chi^2 \beta ^2 - 
2 \sqrt{14}\chi \beta \cos 3\gamma  
+14 \Big )\nonumber\\
&+&
k_3 ~\frac{4}{49\sqrt{5}}~\frac{\beta^3}{(1 + \beta^2)^3} \Biggl( 
   {\chi^3 \beta^3 (2\cos^2{3\gamma}-1)} 
-  \sqrt{14} (3 \chi^2 \beta^2 +14) \cos{3\gamma}
 + 42 \chi \beta   
\Biggr)\Biggr].
\label{eq-ener-large}
\end{eqnarray}
\end{widetext}
Note that the $N$ dependence is eliminated due to the proper $N$
scaling of the different terms in the Hamiltonian and to the use of the
energy per boson.  

The presence of $\cos{3\gamma}$ and
$\cos^2{3\gamma}$ in (\ref{eq-ener-large}) can give rise to triaxiality.
Let us start discussing the possibility of
triaxiality for a simpler situation: an energy surface
independent of $\beta$, but with the $\gamma$ dependence as in
Eq.~(\ref{eq-ener-large}). In that case, one can write the energy surface as 
\begin{equation}
E(\gamma)= a \cos^2{3\gamma} + b \cos{3\gamma} + c
\end{equation}
where we take $a$, $b$, and $c$ as constants for the moment. This
function admits triaxial extrema at 
\begin{equation}
\gamma=(\pm \arccos{(-b/2a)}+ 2 n \pi)/3
\label{gamma}
\end{equation}
if $ | -b/2a | \le 1$, with  $n=0,1,2$. When instead $\mid -b/2a \mid >
1$, the extrema sit at $\gamma=2 \pi n/3$ and
$\gamma=(2 n+1)\pi /3$,  with  $n=0,1,2$, and triaxiality is excluded. 

In the general case, in which the coefficients  $a$ and
$b$ depend on $\beta$, as in the energy functional
 Eq.~(\ref{eq-ener-large}),
the presence of terms depending on $\cos{3\gamma}$ and
$\cos^2{3\gamma}$ can, in principle,
produce triaxial shapes. It is clear that the existence of a
triaxial minimum is linked to the inclusion of the cubic-Q term ($k_3
\neq 0$) in the IBM Hamiltonian.
The argument discussed above could, in principle, be valid even when the coefficients  
$a$ and
$b$ depend on $\beta$, as long as their dependence on the equilibrium value of $\beta$, 
$\beta_0$,
is smooth enough with respect to the control parameters of the Hamiltonian. 
However, for the 
cubic CQF Hamiltonian the latter argument is not so obvious, mainly due to the non
continuous or very ``fast'' dependence of the $\beta$ equilibrium
value in some intervals of the control parameters. 
To shed 
some light on this, one can calculate the
value of $\mid -b/2a \mid$ (see Eq.~(\ref{gamma})), which is the only
necessary quantity to be known for getting the
value of $\gamma$, and it can be written as
\begin{equation}
\left |- \frac{b}{2a} \right |=\left |- \frac{7 \sqrt{70} \chi (1+\beta_0^2) +\sqrt{14}
  k_3 (14+3 \chi^2 \beta_0^2 )}{4 k_3 \chi^3 \beta_0^3 }\right |  \;,
\label{ab}
\end{equation}
where $\beta_0$ is the equilibrium value of
$\beta-$, that is, at the minimum of the energy. To rule out the possible existence
of a triaxial region one has to prove as necessary and sufficient
condition that $|\frac{-b}{2a}|>1$, for any value of the  parameters in the
Hamiltonian. In fact, this is true for $\xi=0$. Note that
Eq.~(\ref{ab}) does not depend explicitly on $\xi$, but its dependence
comes in implicitly through the dependence of the equilibrium
value of $\beta$ on $\xi$, $\chi$, and $k_3$. 

\begin{figure}[hbt]
\begin{center}
\includegraphics[width=5cm]{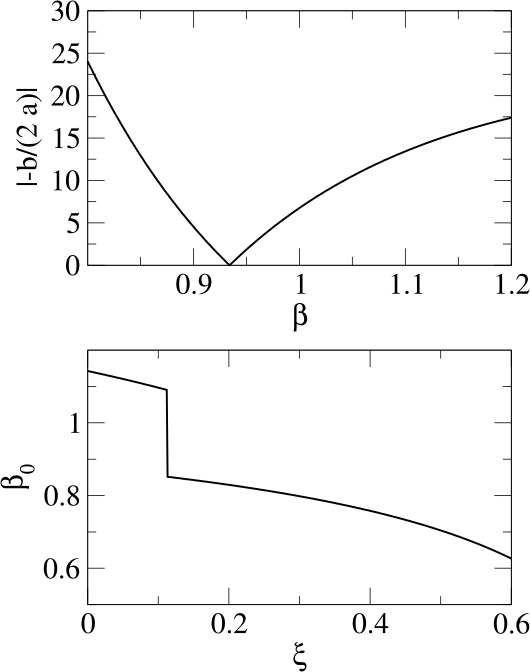}
\caption{Upper part: value of  $|\frac{-b}{2a}|$ (see text) as a function of
  $\beta$ for $k_3=1$ and $\chi=-0.5$. Equilibrium value
  of $\beta$ as a function of $\xi$.}
\label{fig-ab}
\end{center}
\end{figure}

To start with, we can easily calculate the value of $|\frac{-b}{2a}|$
in some limiting situations. In particular, $\lim_{\beta\to
  \infty}|\frac{-b}{2a}|=0$ or $\lim_{|\chi|\to \infty}|\frac{-b}{2a}|=0$ and therefore 
$\gamma=\pi/6$, but these situations are quite unrealistic. 
More interesting limiting situations are: $\lim_{\beta\to 0}|\frac{-b}{2a}|=\infty$,
$\lim_{k_3\to\infty}|\frac{-b}{2a}|>16$ (this case is
  obtained considering that $\beta_0^2 \chi^2$ is always larger than
  zero, remembering that $\beta_0$ in this equation is the value
  at the equilibrium point, taking into account that the larger is $|\chi|$, the larger is 
  $\beta_0$ and, finally, fixing the maximum
  values for $|\chi|$ and $\beta_0$ to $\sqrt{7}/2$ and $1/\sqrt{2}$, respectively), 
$\lim_{k_3\to 0}|\frac{-b}{2a}|=\infty$, and $\lim_{\chi\to
  0}|\frac{-b}{2a}|=\infty$.  We will also prove later on
that $|\frac{-b}{2a}|>1$ for $\xi=0$. 
On the other hand it is also true that the value of
$\beta$ obtained for $\xi=0$ is always larger than the corresponding
one for $\xi\neq 0$ (for the same values of $\chi$ and $k_3$). With these
general ideas in mind, it is easy to see that $|\frac{-b}{2a}|>>1$
except, maybe, for a very 
narrow range of the parameters. Indeed, one can
see that, once the values of the Hamiltonian parameters have been set,
if we treat $\beta$ as a free parameter and not as a equilibirum value, one gets $|\frac{-b}{2a}|<1$
only in a very narrow range of $\beta$.
In particular, it is possible to see, numerically, that
this particular range of $\beta$ is placed
in a region where $\beta$ exhibits a sudden drop. This is
illustrated in Fig. \ref{fig-ab}, where the value of
$|\frac{-b}{2a}|$ is plotted as a function of $\beta$, for $k_3=1$ and
$\chi=-0.5$. It is clearly observed that only around $\beta\approx
0.934$ one gets $|\frac{-b}{2a}|<1$, but this particular value of $\beta$ falls in
a rapidly changing region as it can be 
appreciated in the lower part of the figure, where the equilibrium
value of $\beta$ is represented as a function of $\xi$.

As a conclusion of this qualitative discussion on possible triaxiality
produced by the Hamiltonian (\ref{h23}), it can be said that one
expects that the region supporting triaxial shapes is small, if
any. We will quantify this idea in the following sections.

\subsection{The coordinates}
\begin{figure}
\begin{tabular}{cc}
\includegraphics[width=9cm ,clip=]{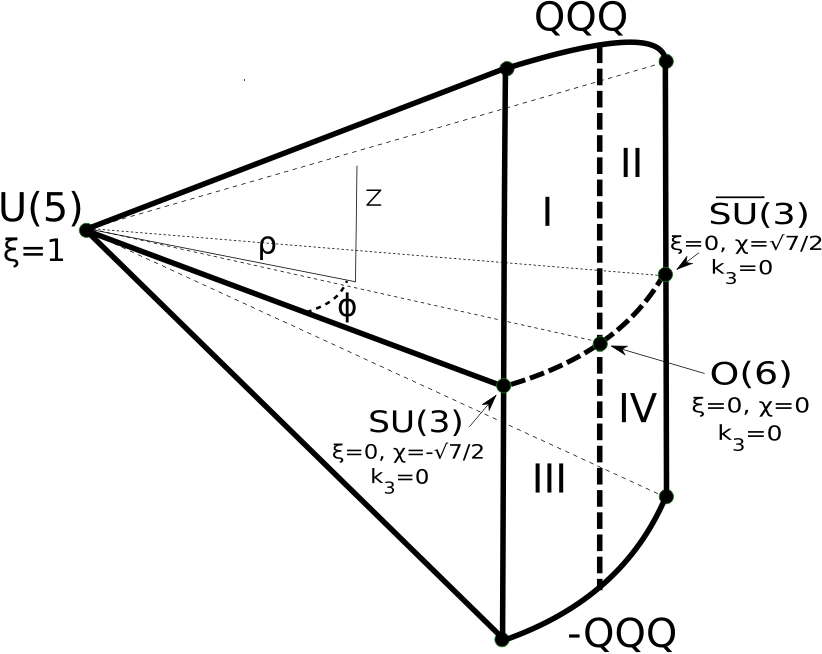} &
\includegraphics[width=5cm,clip=]{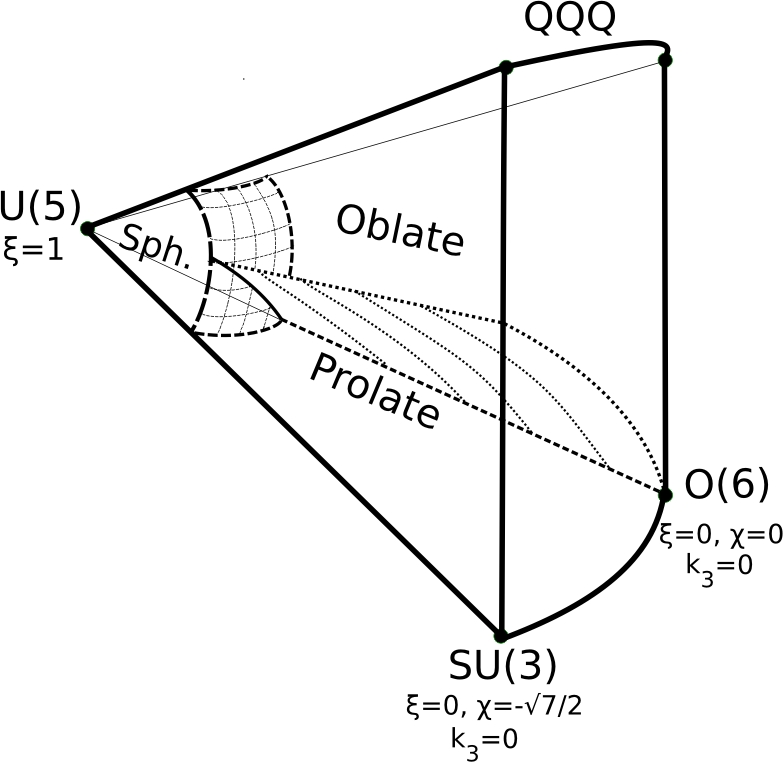}
\end{tabular}
\caption{(Left) Schematic representation of the  CCQH
parameter space and coordinates. Four quadrants are present, of which
only the first one is 
studied in detail as explained in the text. As a reference, the IBM dynamical symmetry 
limits are explicitly shown. (Right)
Schematic phase diagram for the  CCQH, corresponding to quadrant I in the figure on the left.}  
\label{fig-coord}
\end{figure}

The pictorial representation of the phase space of the CCQH is
an extension of the well-known Casten triangle (left plot in
Fig. \ref{fig-coord}). In our case the 
parameter space becomes a tetrahedron, where the horizontal coordinates are
related to the $\xi$ and   $\chi$ control parameters in the Hamiltonian, while the
vertical coordinate is directly connected to the coefficient of the
$\hat{Q}\times \hat{Q}\times \hat{Q}$ 
term. We use the following 
parametrization:
\begin{equation}
\left\{ \rho = 1-\xi ~;~ \phi = - \frac{\pi}{3}~\frac{\chi}{\sqrt{7}} ~;~z=\rho ~k_3  ~ \right\}.
\label{coord}
\end{equation}
In the studies presented below, the value of $\chi$ ranges from $-\sqrt{7}/2$ to $0$ and therefore
the amplitude of the $\phi$ angle is $30^\circ$, $\rho$ ranges from
$0$ to $1$ and $k_3$ is taken as positive. This range of control
parameters allows to get the results for the full model space since the energy surface
(\ref{eq-ener-large}) presents the following  prolate/oblate symmetry:
 $\chi\rightarrow -\chi$, $k_3 \rightarrow -k_3$, $\gamma \rightarrow
 \pi/3- \gamma$. This allows to extend the results obtained in the
 case of $\chi<0$ and $k_3 >0$ to the regions with $\chi>0$ and $k_3<0$.

It is worth noting that due to the form of the Hamiltonian (\ref{h23}) it is not
possible to reach a pure Cubic Q Hamiltonian, except for
$k_3\to\infty$. However in Fig. \ref{fig-coord} the upper part of
the tetrahedron is labeled with $\hat{Q}\times \hat{Q}\times \hat{Q}$ to denote that this limit is reached
for large values of $z=\rho k_3$. Anyway, in this study we will limit
ourselves to moderated large values of $k_3$ (up to 10)
since the cubic Q term is supposed to be a correction to the dominant QQ term.

First, the schematic phase diagram of the model Hamiltonian will be
studied and then some particular relevant regions will be analyzed in
more detail with the aim of singling out  possible regions of triaxiality.

\subsection{The geometry of the phase diagram}

The resulting phase diagram for the  Hamiltonian (\ref{h23}) is
depicted in the right hand side of Fig.~\ref{fig-coord}. 
Although a detailed explanation of the nature of this phase diagram will
be given along this section, we already present the complete diagram here  
to facilitate the following analysis.
The position of the critical surfaces are determined numerically,
except for some particular regions of the phase diagram that can be
obtained analytically. We have explored the parameter space to find out
the position of the critical surfaces, following certain selected paths across 
the parameter space to clearly illustrate where the critical surfaces are placed 
and what is their character. 
This phase diagram is characterized
by the existence of four different phases: spherical ($\rho\lesssim 0.2$, or
$1>\xi\gtrsim 0.8$), prolate and oblate axially deformed shapes
($\rho\gtrsim 0.2$, or
$\xi\lesssim 0.8$) and a very tiny region of triaxial shapes, occurring
between the oblate/prolate separation surface close to the $\chi=\pm
\sqrt{7}/2 $ faces of the tetrahedron. Note that the existence of a
spherical region is limited to moderate values of $k_3$. Due to the
particular form of Hamiltonian (\ref{h23}), very lare values of $k_3$
will transform the spherical region almost in a single point around
the origin.   
The two plotted surfaces correspond
to first order phase transitions except along their intersection line, which is
a second order phase transition line (full line in the right panel of
Fig.~\ref{fig-coord}). As we will explain in detail along this section, the prolate-oblate
first order transition surface becomes a two-fold second order phase
transitions surface for $\chi\approx -\sqrt{7}/2$, although these surfaces are 
extremely close and cannot be distinguished in
Fig.~\ref{fig-coord}. 

The full phase space comprehend four tetrahedra (numbered I - IV) with
a common U(5) vertex,  
shown in the left part of Fig.~\ref{fig-coord}, while in the right
part of the figure only the upper left quarter (I) with positive
values of $k_3$ and negative values of $\chi$ is plotted.  
Apart from the spherical region close to $\xi=1$, that is always
present, the upper right quadrant II ($k_3 >0$ and $\chi >0$) contains
only oblate shapes, while the opposite lower left quadrant III ($k_3
<0$ and $\chi <0$) contains only prolate shapes. Finally, due to the
prolate oblate/symmetry  
$\chi\rightarrow -\chi$, $k_3 \rightarrow -k_3$, $\gamma \rightarrow
\pi/3- \gamma$, the lower right quadrant IV can be obtained by
rotating the first quadrant $180^\circ$ with respect to the 
$U(5)-O(6)$ axis.
In order to illustrate how the phase diagram has been obtained we will
show the results along several selected paths within quadrant I.

\subsubsection{Paths from deformed to spherical shapes}

\begin{figure}
 \begin{tabular}{cc}
 \includegraphics[width=9cm , clip=]{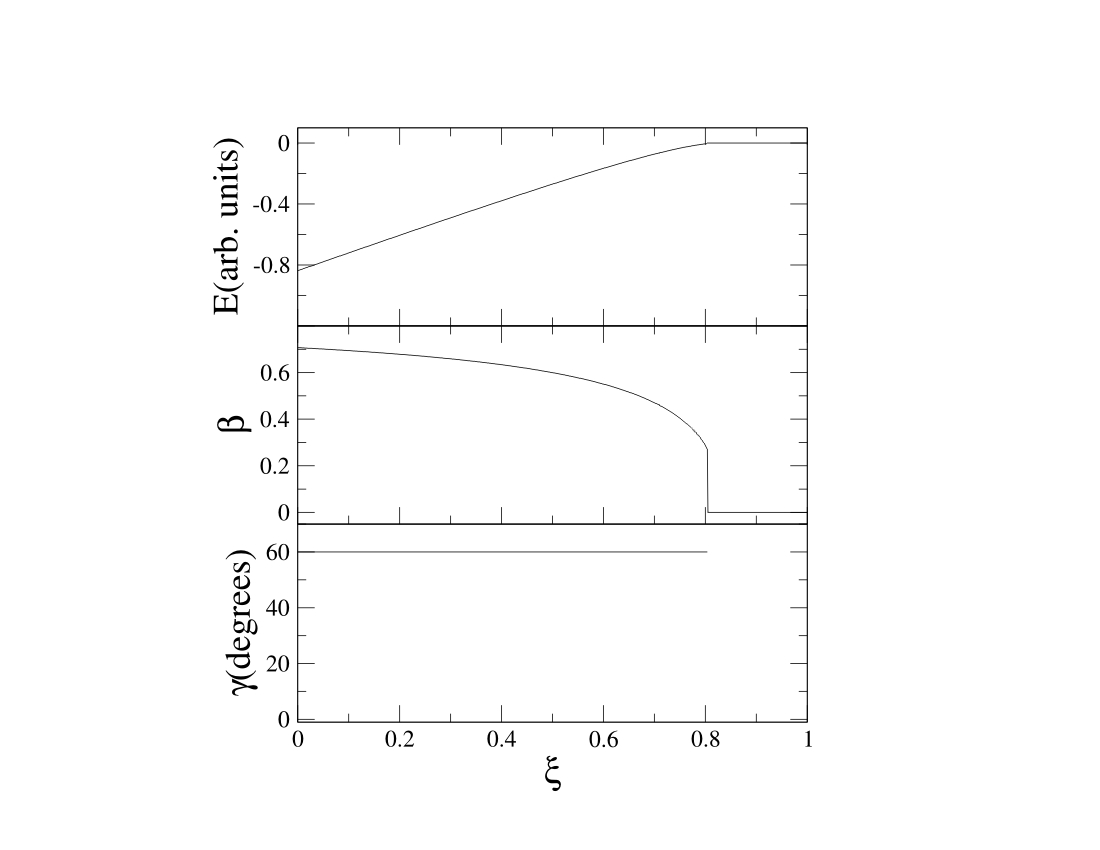} &
 \includegraphics[width=4cm]{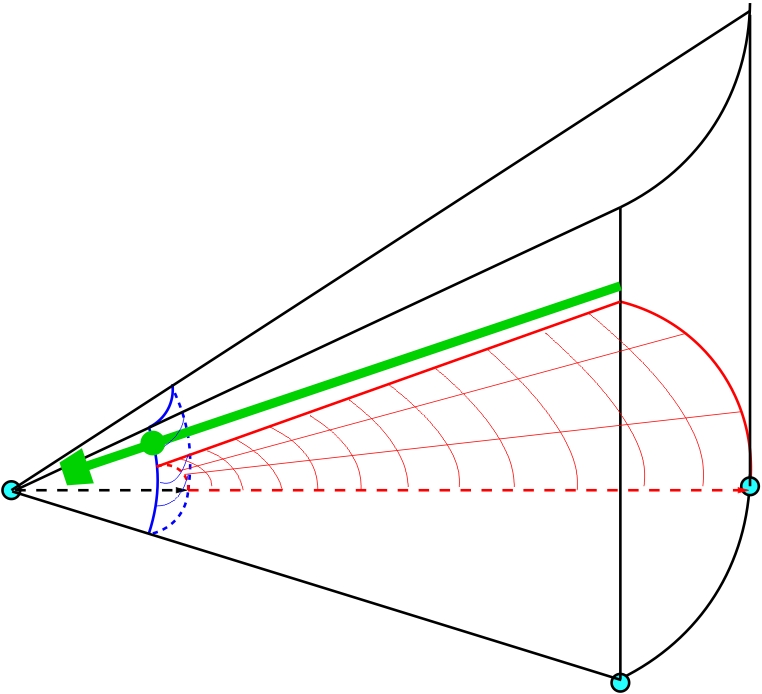} 
 \end{tabular}
\caption{(Color online) Ground state energy and  the equilibrium value of the shape
  variables $\beta$ and $\gamma$ as a function of $\xi$ along the path shown in the right part of the figure, 
for $k_3=4$ and $\chi=-\sqrt{7}/2$. 
Notice that the equilibrium value of $\gamma$ is  undefined in the spherical region.} 
\label{fig-camino2}
\end{figure}

\begin{figure}
\begin{tabular}{cc}
\includegraphics[width=9cm , clip=]{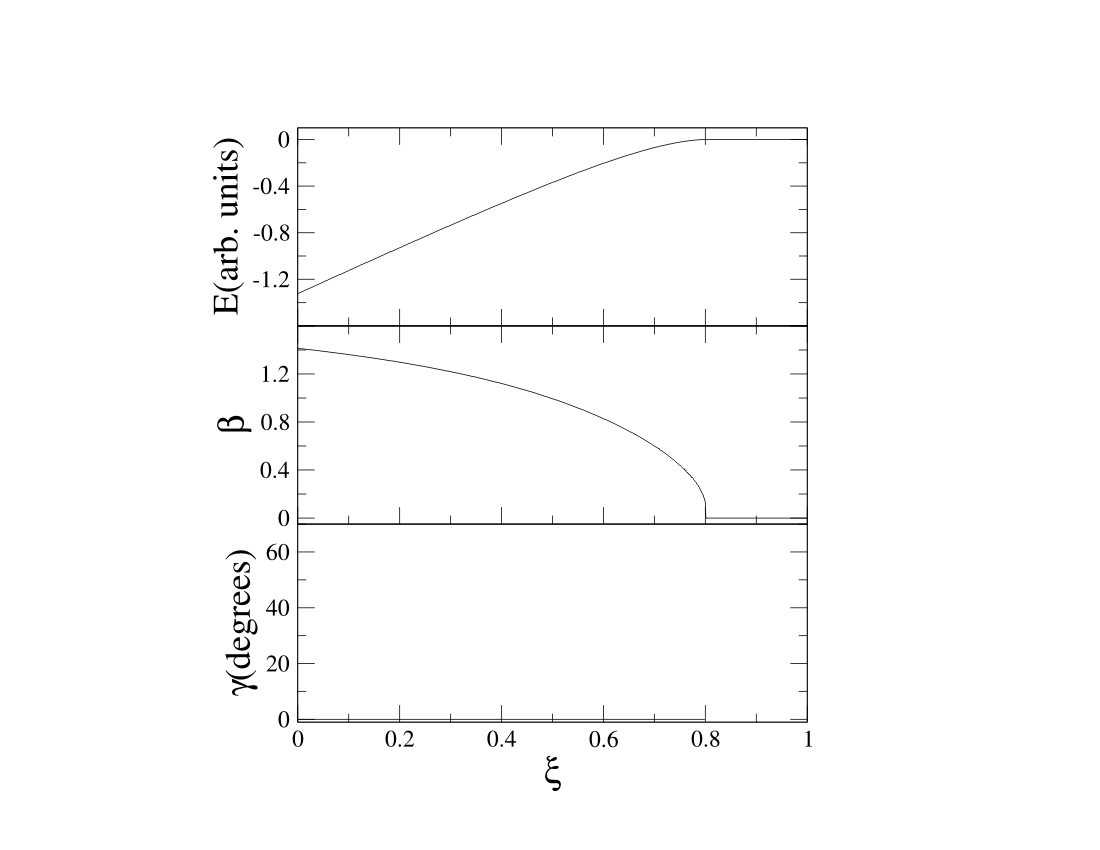}&
\includegraphics[width=4cm]{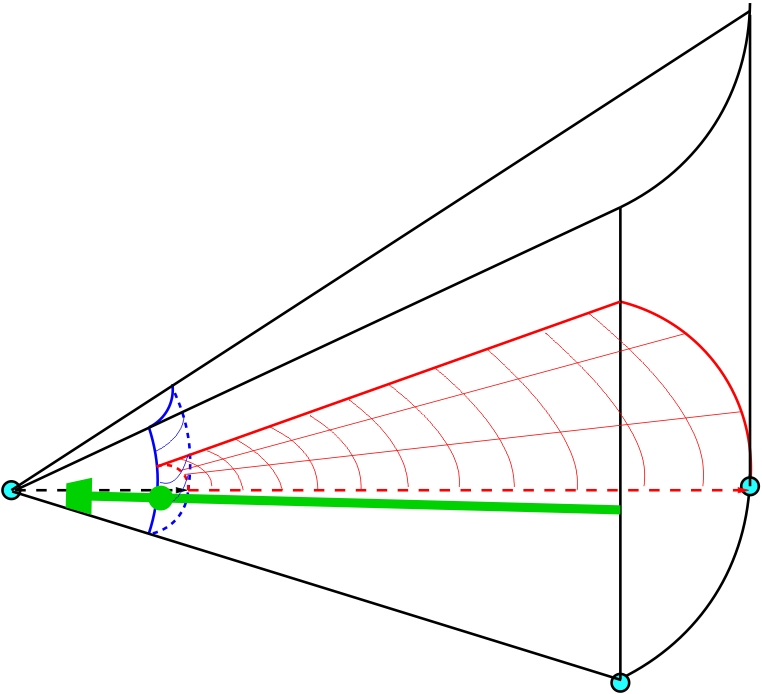}
\end{tabular}
\caption{(Color online) The same as in Fig.~\ref{fig-camino2} but for $k_3=1$ and $\chi=-\sqrt{7}/2$.} 
\label{fig-camino4}
\end{figure}

First, we will explore the separation surface of spherical and
deformed shapes. For that purpose, we start 
selecting $k_3=4$ and $\chi=-\sqrt{7}/2$ and we vary
$\xi$. 
In Fig.~\ref{fig-camino2} we observe a first
order phase transition, where $\beta$ goes from a finite value to
zero and $\gamma$ goes from $60^\circ$ (oblate shape) to undefined (spherical), when entering the
spherical region. Note that the equilibrium value of $\beta$ for
$\xi=0$ is $\beta=1/\sqrt{2}$.  In the
right part of the figure we have displayed the position of the path
(green thick line) inside the phase diagram and also the position of
the phase transition point (green circle). 
It should be noted that U(5) corresponds to $\xi=1$ and the surface SU(3)-O(6)-QQQ to $\xi=0$.  
In the following figures we include plots with similar interpretations.
For $\chi=0$ one can obtain plots similar to Fig.~\ref{fig-camino2}, 
although in this case the transition occurs for a slightly larger
value of $\xi$. 
Now we consider a smaller value of $k_3$. We impose  $k_3=1$ and
$\chi=-\sqrt{7}/2$ and we vary the value of $\xi$. This trajectory is
plotted in Fig.~\ref{fig-camino4} and once more the existence of a
first order phase transition can be noted, although it is not as abrupt as in
the previous case. The value of $\gamma$ passes from zero (prolate
shape) to
undefined in the spherical region. Finally note that $\beta=\sqrt{2}$ for $\xi=0$. 

Now, it is clear that there is a surface that separates spherical and
deformed shapes at around $\xi=0.8$. In order to investigate the
character of this surface,
where the system changes from a spherical to a
deformed shape, it is possible to carry out a Taylor expansion of the
energy around the value $\beta=0$, obtaining:
\begin{eqnarray}
\nonumber
E(\xi,k_3,\chi,\beta,\gamma)&=& (-4 +5\xi)\,\beta^2\\
\nonumber
&-&\frac{4}{35}(1-\xi)\left(2\sqrt{70}
k_3+5\sqrt{14}\chi\right)\,\cos{3\gamma}\,\beta^3\\
&+&\left(-\xi+(1-\xi)\left(8-\frac{24 k_3
  \chi}{7\sqrt{5}}-\frac{2\chi^2}{7}\right)\right)\,\beta^4 
+\Theta(\beta^5) .
\label{e-sp-defor}
\end{eqnarray}
This expression is very convenient because one can easily read off  the order
of the phase transition when crossing the surface. In general, the
presence of a cubic term in $\beta$ implies that the system 
undergoes a first order phase transition \cite{Gilm81}, while its absence guaranties
that the phase transition is of second order. 

From equation (\ref{e-sp-defor}) and the preceding discussion one notices that the
spherical-deformed surface corresponds to a first order phase
transition except when
\begin{equation}
k_3=-\frac{\sqrt{5}}{2} \chi,
\label{k_2o}
\end{equation}
which cancels the $\beta^3$ term and changes the transition type to second order.
This second order phase transition line is the
intersection of the spherical-deformed surface (studied here) and the prolate-oblate
surface (to be studied in the next subsection). Therefore along this
line spherical, prolate and oblate shapes coexist.  

\begin{figure}
\begin{tabular}{cc}
\includegraphics[width=9cm , clip=]{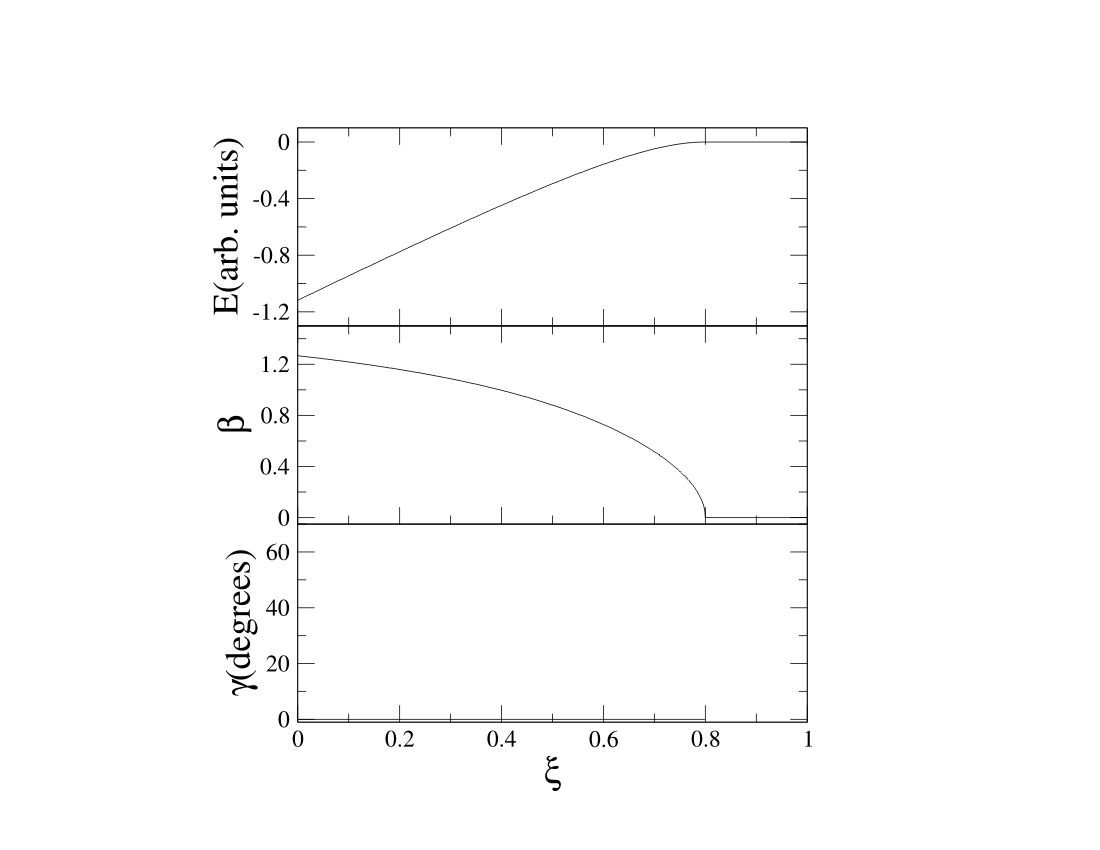}&
\includegraphics[width=4cm]{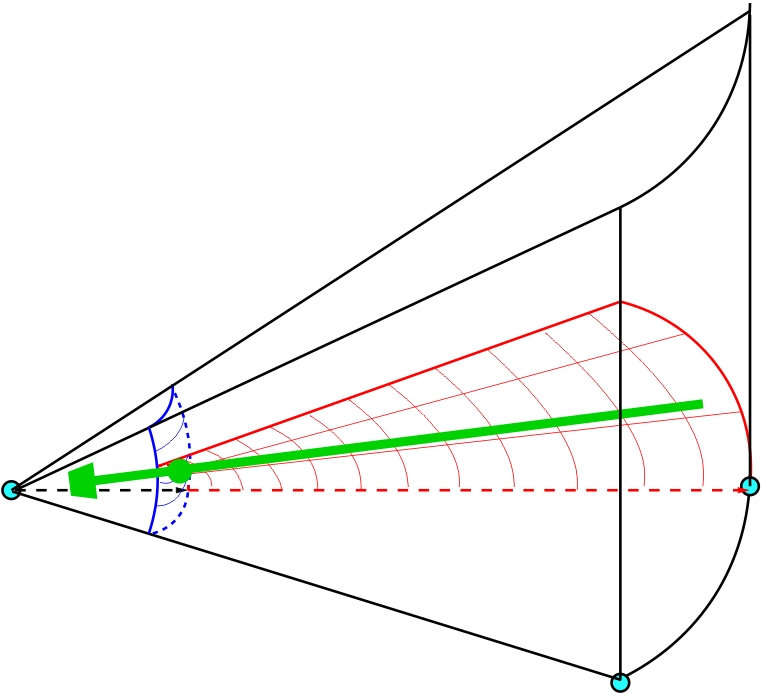}
\end{tabular}
\caption{(Color online) The same as in Fig.~\ref{fig-camino2} or \ref{fig-camino4}
  but for $k_3=1$ and $\chi=-2/\sqrt{5}$ (second order phase transition).}  
\label{fig-camino6}
\end{figure}

In order to show the characteristics of this second order phase
transition line we have followed a path 
that crosses the
spherical-deformed surface precisely through this line. 
This is illustrated in Fig.~\ref{fig-camino6}, where a second
order phase transition appears at $\xi=4/5$.
No discontinuity is observed neither in the order parameter $\beta$,
nor in the energy or in its first derivative (not shown
in the figure). The discontinuity appears in the second derivative of
the energy confirming that the line (\ref{k_2o}) is of the second order type.

With respect to the existence or not of triaxiality in the region
close to spherical shapes, in expression
(\ref{e-sp-defor}) only the $\cos{3\gamma}$ shows up, but the
$\cos^2{3\gamma}$ dependence will appear together with $\beta^6$ terms and therefore
it will generate a value $|\frac{-b}{2a}|>>1$ 
since the coefficient of $\cos{3\gamma}$ (a) is much smaller than the
one of $\cos^2{3\gamma}$ (b) for $\beta < 1$. Therefore,
there is no possibility of triaxiality close to the spherical region. 

Finally, note that for every value of $\xi$, with the exception of
$\xi=1$, it always exists a value of $k_3$ that marks the appearance of an additional 
deformed minimum. This situation is quite different with respect to the one obtained with 
two body Hamiltonians, where a value of $\xi\gtrsim 0.8$ implies a spherical
minimum regardless of the value of $\chi$. In the phase diagram
depicted in the right part of Fig.~\ref{fig-coord} this fact cannot be
noticed because of the moderated values of $k_3$ used along the diagram.

\subsubsection{Paths from deformed prolate to deformed oblate shapes}

\begin{figure}
\begin{tabular}{cc}
\includegraphics[width=9cm]{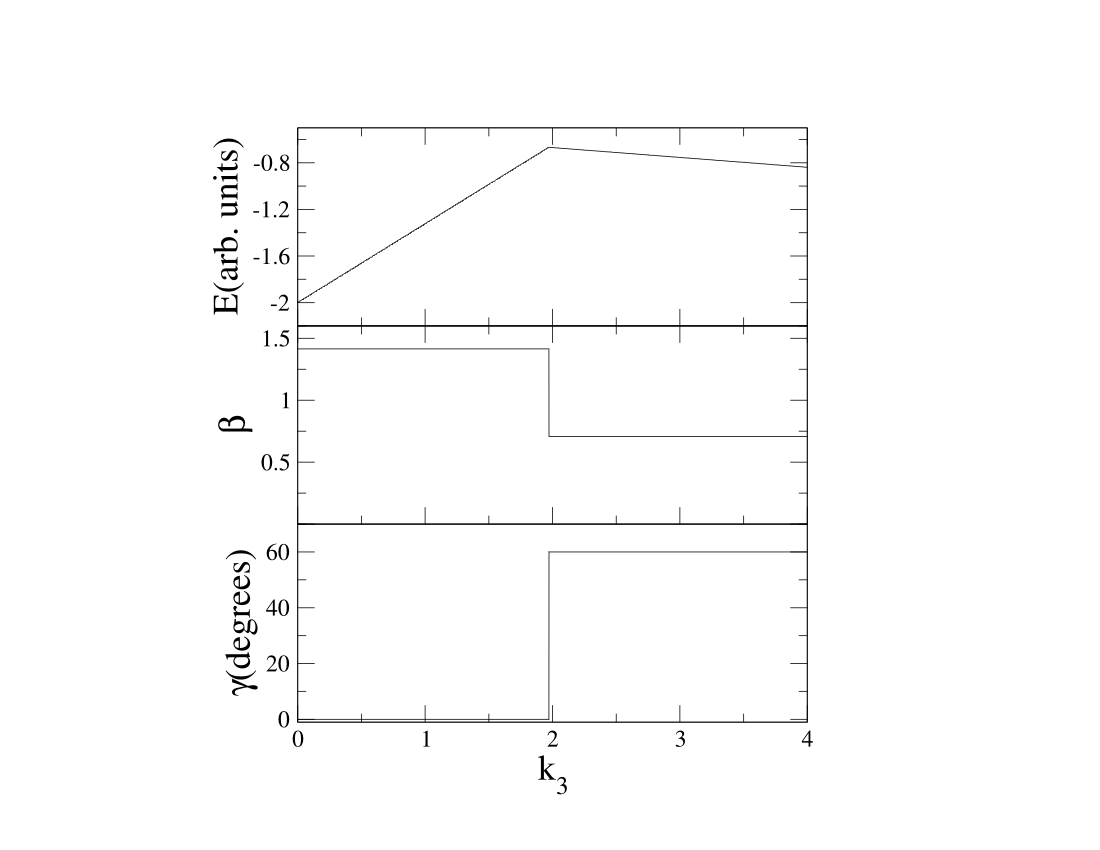}&
\includegraphics[width=4cm]{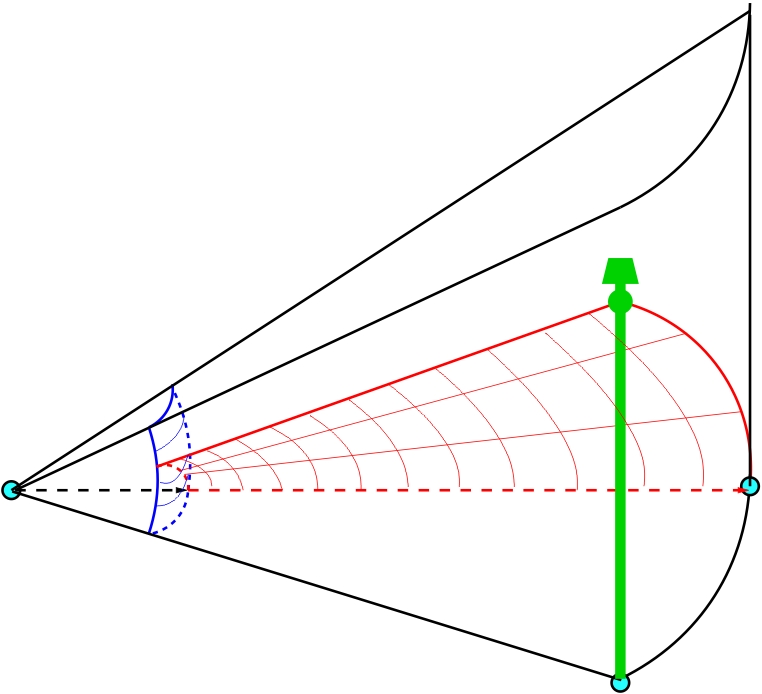}
\end{tabular}
\caption{(Color online) Ground state energy, equilibrium shape variables $\beta$ and
  $\gamma$ as a function of $k_3$ for $\xi=0$ and $\chi=-\sqrt{7}/2$.} 
\label{fig-camino1}
\end{figure}  
Now, we explore the separation surface between prolate and oblate
shapes, starting from the $\xi=0$ surface and then generalizing to
$\xi \ne 0$.

\paragraph{ The $\xi=0$ case: interplay of $\hat{Q}\cdot \hat{Q}$
  and  $(\hat{Q}\times \hat{Q}\times \hat{Q})^{(0)}$\\ \\} 

In this subsection the  $\xi=0$ ($\rho=1$) surface (SU(3)-O(6)-QQQ) is studied.
For pedagogical reasons we start our study by setting
$\chi=-\sqrt{7}/2$ (the vertical line between SU(3) and QQQ in
Fig.~\ref{fig-coord}) and later on we will generalize the outcome to arbitrary
values of $\chi$. For the case,  $\xi=0, \chi=-\sqrt{7}/2$, 
the extrema of the $-\hat{Q}\cdot \hat{Q}/N^2$ term
in the large N limit are listed in Table \ref{tab-QQ-ext}. In this Table 
the signs of the second derivative of the energy with
respect to both shape parameters, $\beta$ and $\gamma$, and the character of the extrema 
(M for maximum, m for minimum and s.p. for saddle point) are given. 
To avoid equivalent shapes related with the symmetries of the P.E.S.

we impose the constraints $\beta\ge 0$ and $0 \le \gamma \le \pi/3$. 
Summarizing the information in Table \ref{tab-QQ-ext},  
the surface obtained from the $-\hat{Q}\cdot \hat{Q}/N^2$ term
presents: a spherical maximum with no dependence on $\gamma$, a
prolate minimum for $\beta=\sqrt{2}$, a saddle point in
the oblate side for $\beta=1/\sqrt{2}$ and an oblate maximum for
$\beta=2\sqrt{2}$. Note that at the extrema there are no off-diagonal
terms in the Hessian matrix.
\begin{table}
\begin{tabular}{ccccc}
\hline
\hline
  $~~~~\beta~~~~$      & $~~~~\gamma~~~~$       & $\partial^2 E/\partial \beta^2$ &
  $\partial^2 E/\partial \gamma^2$   &   character\\ 
\hline
  0            & $-$      & $0$ & $-$    &   M\\
  $1/\sqrt{2}$ & $\pi/3$  & $+$ & $-$    &   sp\\
  $\sqrt{2}$   & $0$      & $+$ & $+$    &   m\\
  $2\sqrt{2}$  & $\pi/3$  & $-$ & $-$    &   M\\
\hline
\hline
\end{tabular}
\caption{Extrema of $-\hat{Q}\cdot \hat{Q}/N^2$, for $\chi=-\sqrt{7}/2$, along with their 
character. The sign of the 
second derivatives with respect to $\beta$ and $\gamma$ are also given
(see text).}
\label{tab-QQ-ext} 
\end{table}   
A similar study can be done for $-(\hat{Q}\times \hat{Q}\times
\hat{Q})^{(0)}/N^3$. The characters of the extrema for the surface produced by this term
are listed in 
Table \ref{tab-QQQ-ext}. In this case there are a spherical
maximum, an oblate minimum at $\beta=1/\sqrt{2}$,
a prolate maximum at $\beta=\sqrt{2}$  and, finally, an oblate saddle
point at $\beta=2\sqrt{2}$. Moreover there is also a minimum for
$\beta\to\infty$ and $\gamma=\pi/6$.
Note that, again, at the extrema there are no off-diagonal
terms in the Hessian matrix.  It is worth noting that the extrema for
both,  $-\hat{Q}\cdot \hat{Q}/N^2$ and $-(\hat{Q}\times \hat{Q}\times
\hat{Q})^{(0)}/N^3$ coincide, although their character, either minimum
or maximum, can be different. 
\begin{table}
\begin{tabular}{ccccc}
\hline
\hline
  $~~~~\beta~~~~$  & $~~~~\gamma~~~~$  & $\partial^2 E/\partial \beta^2$ & $\partial^2 E/\partial \gamma^2$   &   character\\
\hline
  0            & $-$             & $0$ & $-$    &   sp\\
  $1/\sqrt{2}$ & $\pi/3$         & $+$ & $+$    &   m\\
  $\sqrt{2}$   & $0$             & $-$ & $-$    &   M \\
  $2\sqrt{2}$  & $\pi/3$         & $0$ & $0$    &   s.p.\\
\hline
\hline
\end{tabular}
\caption{Same as Table \ref{tab-QQ-ext} but for $-(\hat{Q}\times
  \hat{Q}\times \hat{Q})^{(0)}/N^3$. }
\label{tab-QQQ-ext}
\end{table}   

A Hamiltonian built upon a linear combination of $-\hat{Q}\cdot \hat{Q}$ and
$-(\hat{Q}\times \hat{Q}\times \hat{Q})^{(0)}$ is expected to possess
eigenstates with large quadrupole moments
and small fluctuations around their equilibrium values. This kind of
state is  eigenstate of the quadrupole operator, as well as 
of both, $\hat{Q}\cdot \hat{Q}$ and $(\hat{Q}\times \hat{Q}\times
\hat{Q})^{(0)}$ \cite{GK,Jolo04}. 
This is the reason why the extrema points of both
parts of the Hamiltonian under study coincide  \cite{Jolo04}. 

In general, a linear combination of $-\hat{Q}\cdot \hat{Q}$ and
$-(\hat{Q}\times \hat{Q}\times \hat{Q})^{(0)}$ always generates a
spherical maximum for $\beta=0$, 
an oblate maximum or saddle point at $\beta=2\sqrt{2}$ and 
a competition between two minima, one
prolate at $\beta=\sqrt{2}$ and the other one oblate  
at $\beta=1/\sqrt{2}$. 
The above situation corresponds to
a first order phase transition and it excludes the presence of a
triaxial minimum.  

Fig.~\ref{fig-camino1} gives the evolution of the
ground state energy and the equilibrium value of $\beta$ and
$\gamma$ for the path along the line $\chi=-\sqrt{7}/2$ and
$\xi=0$. This figure shows the typical ingredients of a first order
phase transition. There is a discontinuity in the first derivative of
the energy and a discontinuity in the value of $\beta$ and
$\gamma$. For low values of $k_3$ the minimum corresponds to
$\beta=\sqrt{2}$ and $\gamma=0^\circ$, while for large values to
$\beta=1/\sqrt{2}$ and $\gamma=60^\circ$. The critical value of the
control parameter corresponds to $k_3=\sqrt{35}/{3}$. This figure is
similar to Fig.~2 appearing in Ref.~\cite{Aria04} although there the
phase transition is second order and $\gamma$ goes from $0^\circ$ to $60^\circ$
in a smooth way and therefore a broad region of triaxiality exists.

\begin{figure}
\begin{tabular}{cc}
\includegraphics[width=6cm]{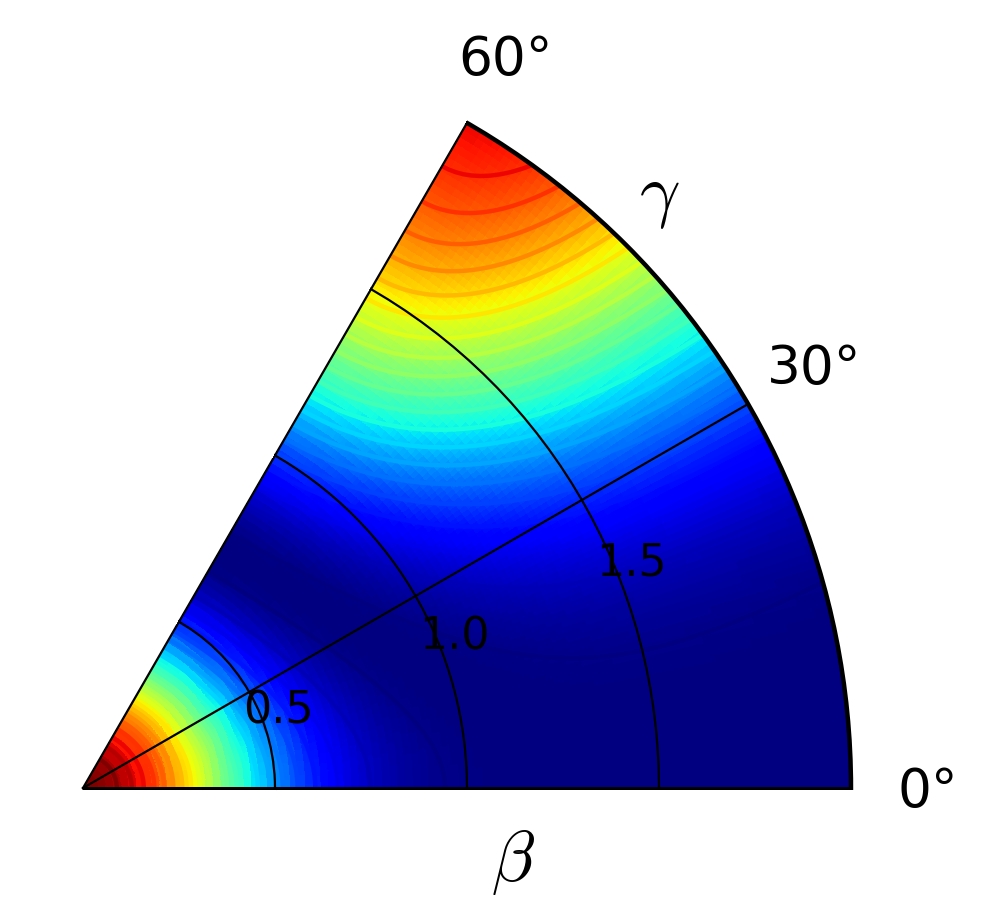}&
\includegraphics[width=5cm]{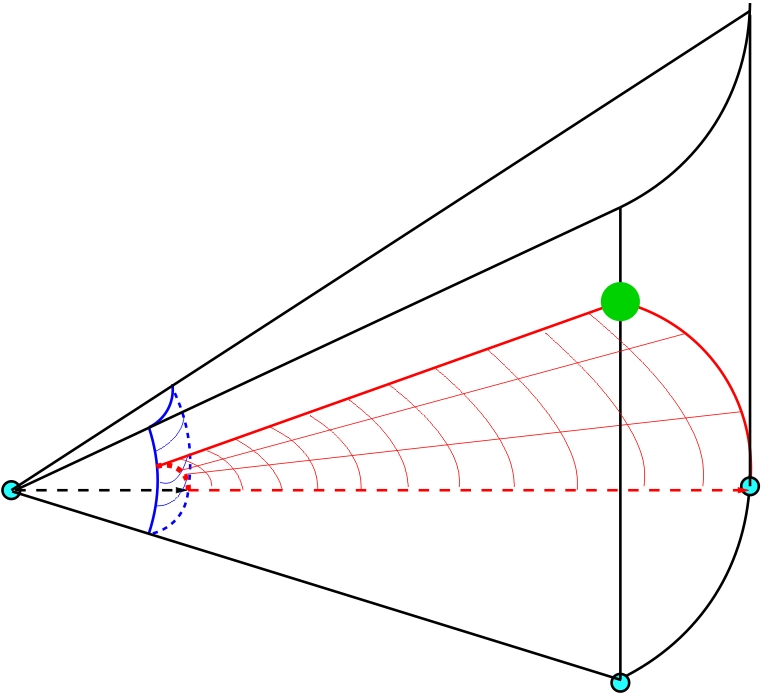}
\end{tabular}
\caption{(Color online) Energy surface for $-\hat{Q}\cdot
  \hat{Q}/N^2-\frac{\sqrt{35}}{3} (\hat{Q}\times 
  \hat{Q}\times \hat{Q})^{(0)}/N^3$ with  $\chi=-\sqrt{7}/2$,
  corresponding to the green dot of the right part of the figure. Note
  the existence of a prolate and an oblate degenerated minima.} 
\label{fig-k35-3} 
\end{figure}

An interesting case happens for 
$k_3=\frac{\sqrt{35}}3$,  for which the prolate minimum ($\beta=\sqrt
2$) flattens (becoming a maximum for larger values),
{\it i.e.}~$\partial^2 E/\partial \beta^2=\partial^2 E/\partial
\gamma^2=0$. Additionally, the flat prolate and 
the oblate minima become
degenerate with energy $E=-\frac{2}{3}$, being also degenerate
with the minimum at $\beta\rightarrow\infty$, $\gamma=\pi/6$. This is illustrated in Fig.~\ref{fig-k35-3}.
For $k_3<\frac{\sqrt{35}}{3}$ the only minimum is the prolate one,
that becomes flat for $k_3=\frac{\sqrt{35}}{3}$. For
$k_3>\frac{\sqrt{35}}{3}$, the only 
minimum is the oblate one. Therefore, at 
$k_3=\frac{\sqrt{35}}{3}$ the spinodal and the antispinodal points
coincide and also are degenerate (first order phase transition point). 

\begin{figure}[hbt]
\begin{center}
\includegraphics[width=5cm]{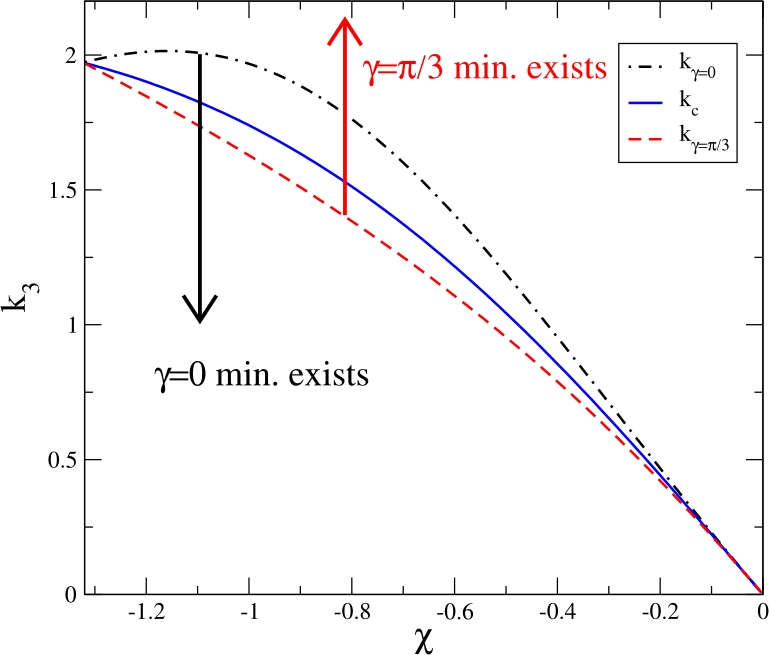}
\caption{(Color online) Value of the spinodal ($\gamma=\pi/3$), antispinodal ($\gamma=0$) and critical value lines of
  $k_3$ as a function of $\chi$ for $\xi=0$.}
\label{fig-k-su3-o6-tr}
\end{center}
\end{figure}

It is possible to extend the preceding study to a generic value of
$\chi$. Without loss of generality, we choose $\chi<0$. In this case
the position of the possible minima are given in Table \ref{tab-QQ-QQQ}. 
\begin{table}
\begin{tabular}{ccc}
\hline
\hline
  $\beta$      & ~~~~~~~~~~$\gamma$~~~~~~~~~     &     \\
\hline
& & \\
$\displaystyle{\frac{\chi}{\sqrt{14}}+\frac{\sqrt{14+\chi^2}}{\sqrt{14}}}$ & $\pi/3$ \\
& & \\
$\displaystyle{-\frac{\chi}{\sqrt{14}}+\frac{\sqrt{14+\chi^2}}{\sqrt{14}}}$& 0 \\
& & \\
\hline
\hline
\end{tabular}
\caption{Position of the possible minima for a linear combination 
$-\hat{Q}\cdot \hat{Q}/N^2-k_3(\hat{Q}\times \hat{Q}\times \hat{Q})^{(0)}/N^3$, with a generic
value of $\chi<0$.}
\label{tab-QQ-QQQ}
\end{table}   
Again there is a competition between the prolate minimum, coming from the
$-\hat{Q}\cdot \hat{Q}/N^2$ term an the oblate one, generated with the 
$-(\hat{Q}\times \hat{Q}\times \hat{Q})^{(0)}/N^3$ term. In this case there exists a
coexistence region, but again there is no room for triaxiality. In
particular, the position of spinodal ($\gamma=\pi/3$) and
antispinodal ($\gamma=0$) points are given by 
\begin{widetext}
\begin{equation} 
k_3 \left(\left.\partial^2 E/\partial \gamma^2\right|_{\gamma=\left\{{\pi/3
  \atop 0}\right\}}=0\right)= -\frac{49 \sqrt{5} 
  \chi  \left(\chi ^2\pm\chi \sqrt{\chi ^2+14} 
    +14\right)}{4 \chi ^6 \pm 4 \chi ^5 \sqrt{\chi ^2+14} +63 \chi ^4
  \pm 35  \chi ^3 \sqrt{\chi ^2+14} +147 \chi ^2+686} ,
\end{equation}   
\end{widetext}
the upper (lower) sign corresponds to $\gamma=\pi/3$ (0).
The value of $k_3$ where the two minima become degenerate (critical
line) reads
\begin{equation} 
k_{3c}=-\frac{7 \sqrt{5} \chi }{2 \chi ^2+7}.
\label{k_c}
\end{equation}   
In Fig.~\ref{fig-k-su3-o6-tr} the spinodal, the antispinodal and the critical
values of $k_3$ are represented in the $(k_3,\chi)$ plane for
$\xi=0$. This figure clearly shows 
the existence of a coexistence region and the absence of a triaxial
region in the $\xi=0$ surface.  

Additional extrema to the ones given in the above mentioned table can appear which do not correspond to 
minima.  

\paragraph{ The $\xi \ne 0$ case: The prolate-oblate-triaxial critical
  surface\\ \\}

The analytical arguments used in the preceding discussion are no longer valid
when the $U(5)$ term gives a contribution to the energy surface,
{\it i.e.}~$\xi\neq 0$, however they suggest that the
parameter space region that gives rise to a triaxial minimum should be rather
small. It is not possible to obtain analytically the expression of the
equilibrium value of  $\gamma$ for a general energy surface. 

\begin{figure}[!t]
\begin{tabular}{cc}
\includegraphics[width=9cm]{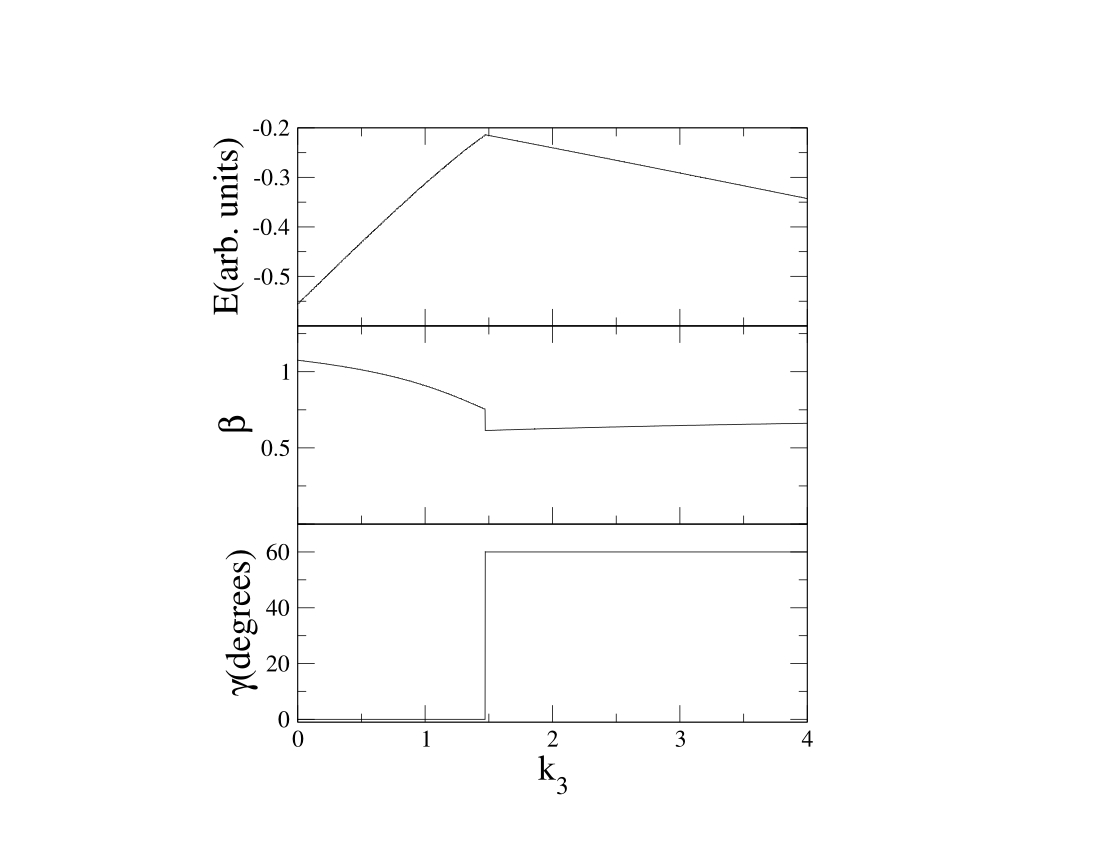}&
\includegraphics[width=4cm]{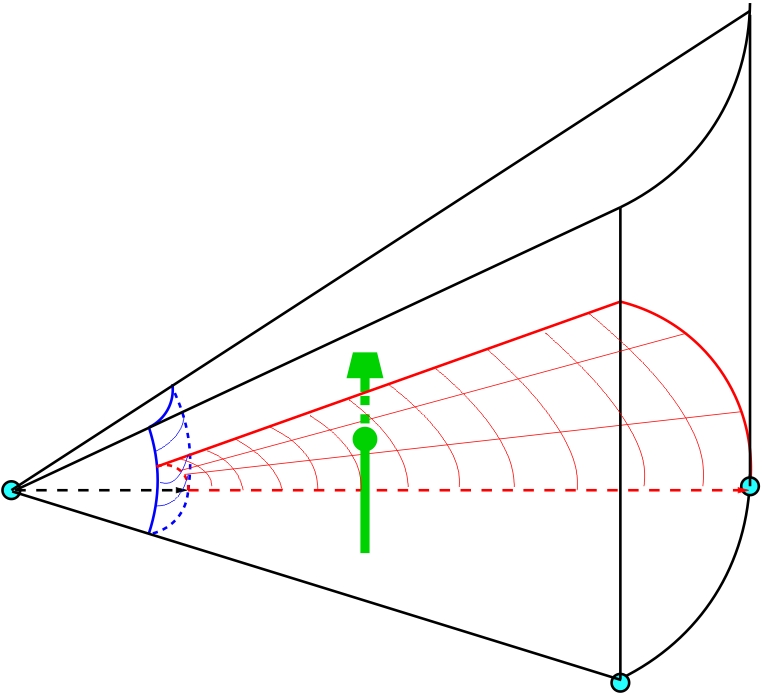} \\
\end{tabular}
\caption{(Color online)  Ground state energy and equilibrium value of the shape variables $\beta$ and
  $\gamma$ as a function of $k_3$, for $\xi=0.5$ and $\chi=-1$ for the path shown on the right.} 
\label{fig-camino8}
\end{figure}

To study numerically the deformed region for this
particular Hamiltonian we calculate the value of the
energy, $\beta$ and $\gamma$ along some particular lines that go
through the phase diagram and cross the prolate-oblate separation
surface. In Fig.~\ref{fig-camino8} we select a particular path with $\xi=0.5$ and
$\chi=-1$ with $k_3$ ranging from $0$ to $4$.    
In this figure the presence of a first order phase transition shows up 
because there exists a clear discontinuity in the values of
$\beta$ and $\gamma$, when crossing the prolate-oblate
surface. Therefore there is a jump from the prolate to the oblate
minimum at the point where they become degenerate and the triaxiality is
forbidden. The same happens for other values of $\xi$ and $\chi$.

\begin{figure}[!t]
\includegraphics[width=10cm, clip= ]{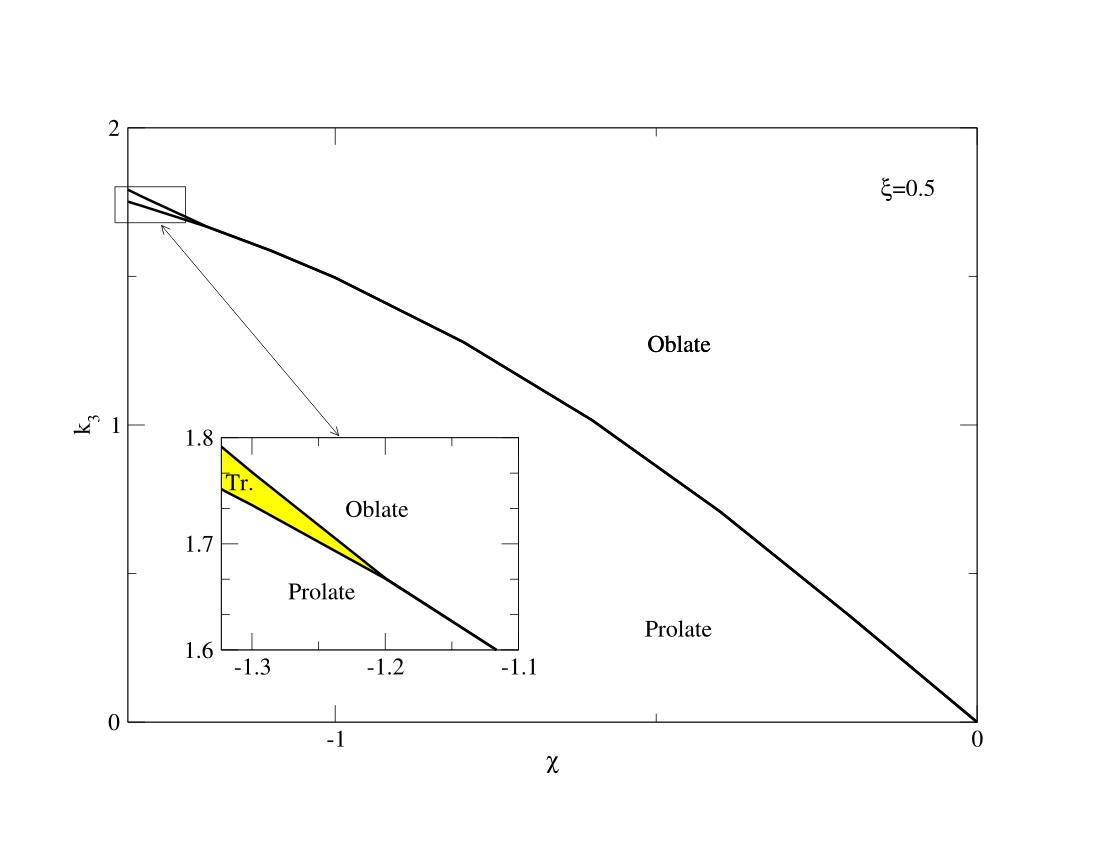}
\caption{(Color online) Section of the three-dimensional parameter space with
  $\xi=0.5$ and $N \rightarrow \infty$. The yellow inset shows the tiny
  triaxial region (see text) that is found at $\chi \approx -\sqrt{7}/2$.} 
\label{kchi}
\end{figure}

However a more detailed inspection shows that there exists a very
tiny region of triaxiality around 
$\chi\approx -\sqrt{7}/2$,  for $\xi<4/5$.
This is clearly illustrated in Fig.~\ref{kchi}, where the lines that separate the prolate,
oblate and triaxial regions are plotted for a value of
$\xi=0.5$. Note that the triaxial region occurs 
for $\chi$ close to the limiting value of $-\sqrt{7}/2$.  At this limit 
$k_3 \in [1.752 - 1.791]$. The lower part of
the figure corresponds to prolate minima, while the upper part to
oblate minima. The box on the vertical left side shows the blowup of this region:
 there is indeed a small range of values where it is possible to find
 triaxial minima by numerical 
procedure. The more one moves away from the value $\chi = 
-\frac{\sqrt{7}}{2}$, the narrower this range becomes, due to the
delicate interplay between the quadratic and cubic terms. 

\begin{figure*}[!t]
\begin{center}
\begin{tabular}{ccccc}
\includegraphics[width=0.19\textwidth,bb= 0 0 315 315]{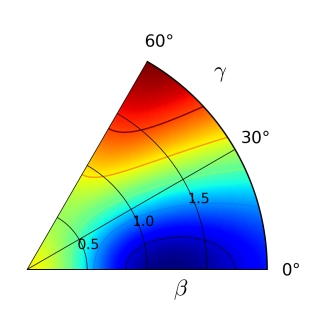} &
\includegraphics[width=0.19\textwidth,bb= 0 0 315 315]{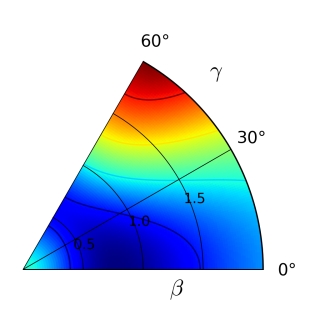} &
\includegraphics[width=0.19\textwidth,bb= 0 0 315 315]{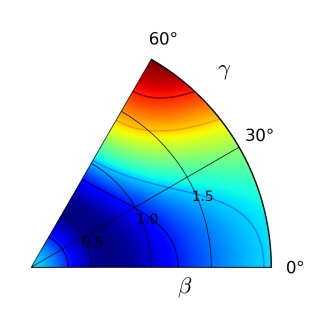} &
\includegraphics[width=0.19\textwidth,bb= 0 0 315 315]{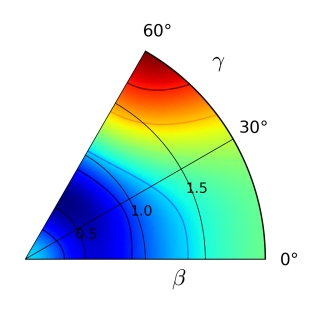} &
\includegraphics[width=0.19\textwidth,bb= 0 0 315 315]{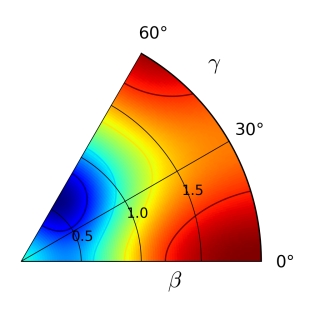} \\
$k_3=0.0$ &$k_3=1.5$ &$k_3=1.77$ &$k_3=2.0$ &$k_3=3.0$ 
\end{tabular}
\caption{(Color online) Potential energy surfaces for the  CCQH
 for $\chi=-\sqrt{7}/2$ and $\xi=0.5$, with different
  values of $k_3$ indicated in the figure. The observed minima range 
from an axially deformed prolate minimum for $k_3=0.0$ and $k_3=1.5$ 
to an oblate one for $k_3=2.0$ and $k_3=3.0$ , passing through the
  triaxial region. The triaxial minimum when $k_3=1.77$ is not
  apparent from this figure, because it is very shallow (see next
  figure).}
\label{vert-ax}
\end{center}
\end{figure*}

\begin{figure}[!t]
\includegraphics[width=5cm]{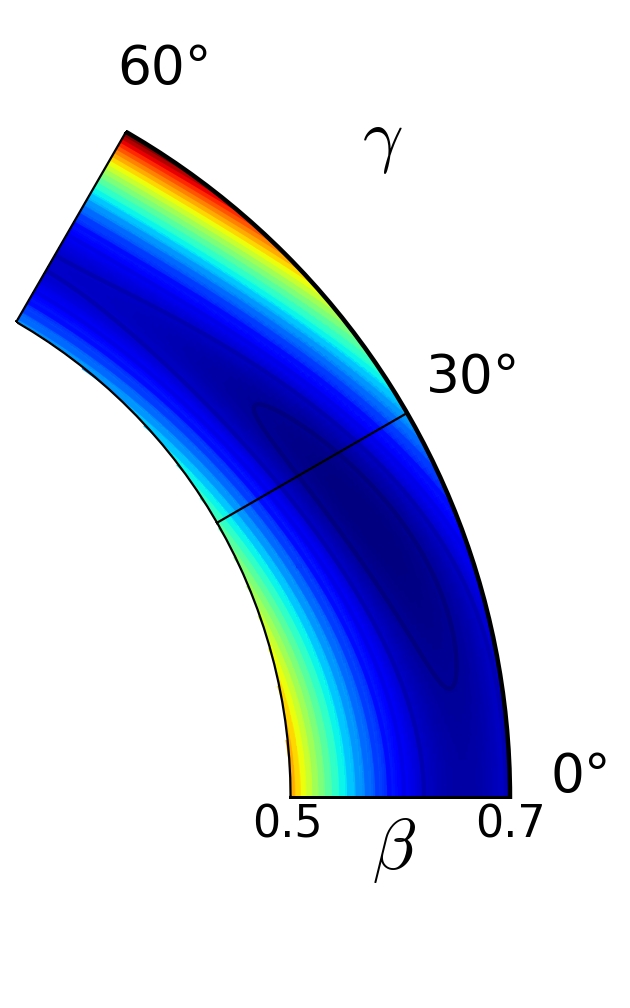}
\caption{(Color online) Closeup of the potential energy surface of Eq. (\ref{eq-ener-large}) 
with  $\chi=-\sqrt{7}/2$, $\xi=0.5$ and $k_3=1.77$, clearly 
 showing a shallow triaxial minimum at $\beta \sim 0.62 $ and
  $\gamma \sim 23.4^\circ$.  One can prove, with similar plots, that the prolate minimum smoothly
shifts through the 
triaxial region and finally becomes oblate.} 
\label{closeup}
\end{figure}

We can see in a clearer way the onset of a triaxial minimum in
Fig.~\ref{vert-ax}. Here we have plotted several
potential energy surfaces corresponding to points along the left
vertical axis of Fig. \ref{kchi} with $\chi = -
\frac{\sqrt{7}}{2}$ and $\xi=0.5$. The minimum is clearly
prolate for small values of $k_3$ and it starts to widen as one
approaches the region $k_3\sim 1.7$. From approximately
$1.752$ to $1.791$ the minimum is triaxial, but very shallow and
therefore invisible at the present scale of $(\beta, \gamma)$. We show
in Fig.~\ref{closeup}, for completeness sake, the close up of
the minimum for $k_3=1.77$. Increasing again the value of $k_3$
leads to an axially deformed oblate minimum.  
It is worth noting that the change through the triaxial region is extremely
swift and some care must be taken in the minimization. 
One can appreciate the extremely shallow character of the triaxial minimum
that extends from the axially prolate till the oblate minimum, very
much in a similar way to Fig.~\ref{fig-k35-3} where prolate and
oblate minima are degenerate and no triaxial minimum exists.  
The existence of the triaxial region is confirmed by inspecting contour plots of the
potential energy surface along a suitable trajectory of the parameter
space, rather than just looking at the plots of $\beta$ and $\gamma$ as a function of the 
parameters
$\xi$, $\chi$ and $k_3$.

\begin{figure}[!t]
\includegraphics[width=10cm, clip= ]{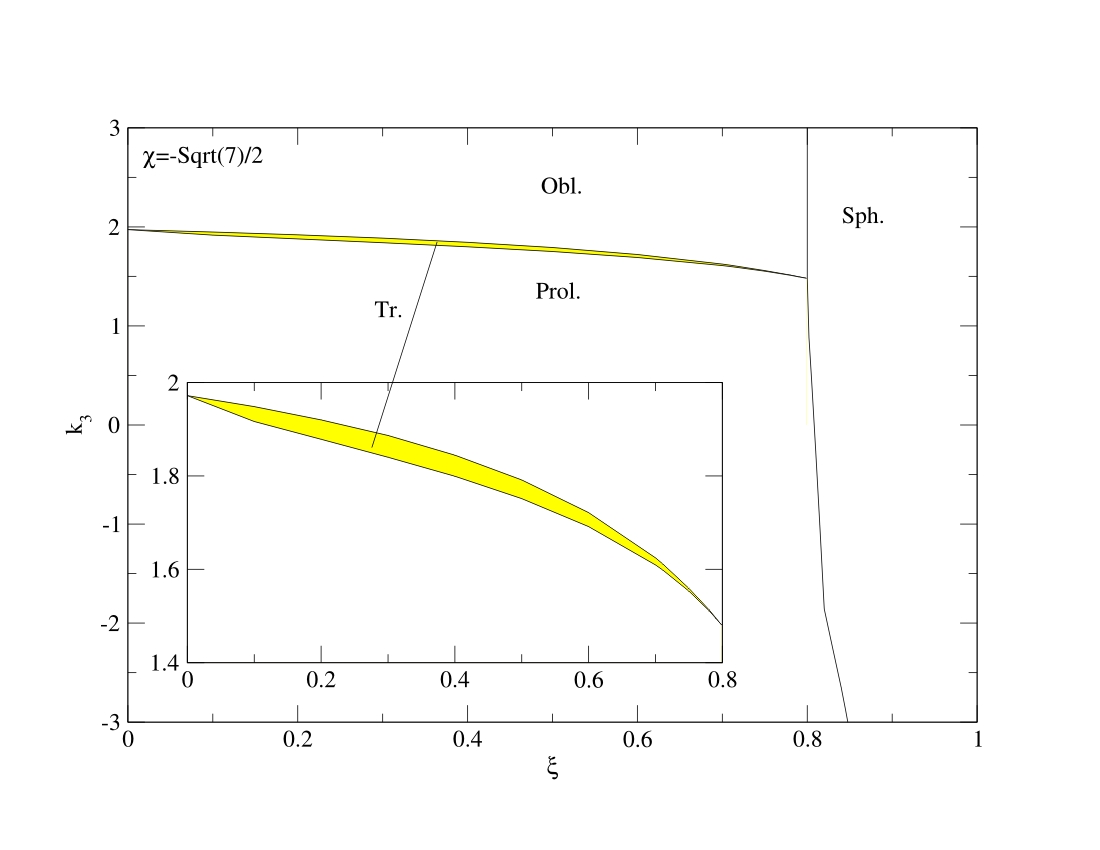}
\caption{(Color online) Section of the three-dimensional parameter space with
  $\chi=-\frac{\sqrt{7}}{2} $ as a function of $\xi$ and $k_3$. The
  inset shows the triaxial region in yellow on a finer vertical scale (see text).} 
\label{kcsi}
\end{figure}

Finally we present in Fig.~\ref{kcsi} a cut of the triaxial
region as a function of $\xi$ and $k_3$ for a constant value of $\chi=-\sqrt{7}/2$. 
A very small triaxial region is always found, except for $\xi=0$,
between the lower prolate region and the upper 
oblate one (see inset of Fig.~\ref{kcsi}) and terminates in
correspondence with the spherical phase that occurs for high values of
$\xi$.  
In a preceding subsection we have discussed that the point connecting with the
spherical phase is indeed a single point as well as the one at $\xi=0$.

A final question should be answered: what is the order of the phase
transition developed in crossing the prolate-triaxial-oblate surface?
In the region of triaxiality the surface evolves into two different sheets
where second order phase transitions exist. As far as we move from
$\chi=-\sqrt{7}/2$ this two-fold surface collapses in a single one and
the phase transition becomes of the first order.

\begin{figure}
\begin{tabular}{cc}
\includegraphics[width=6cm, clip=]{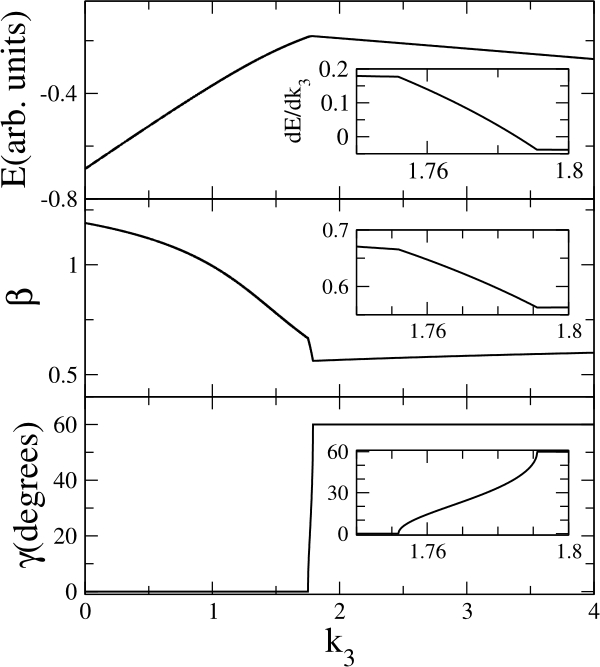}&
\includegraphics[width=4cm]{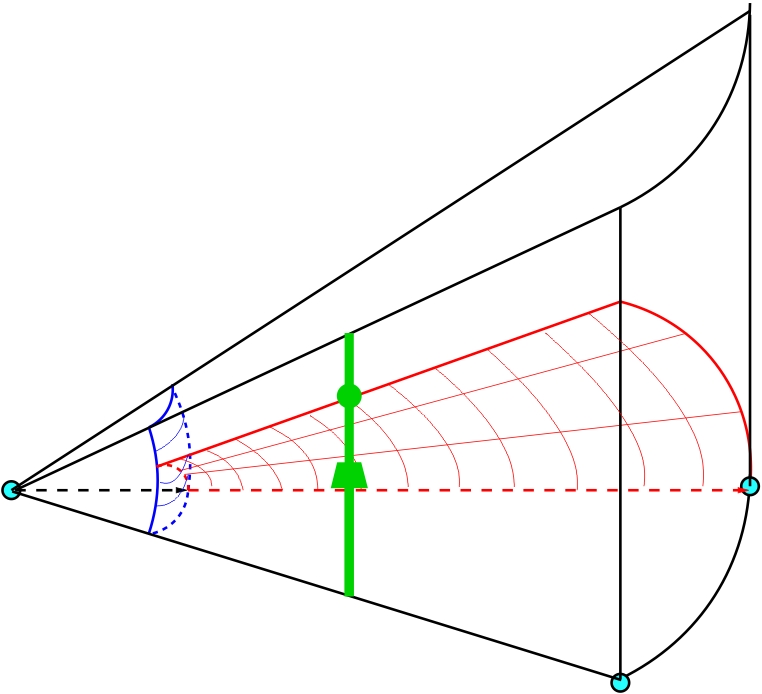}
\end{tabular}
\caption{(Color online) Ground state energy, shape variables $\beta$ and
  $\gamma$ as a function of $k_3$ for $\xi=0.5$ and $\chi=-\sqrt{7}/2$ for the path shown on the right.} 
\label{fig-camino1b}
\end{figure}

Since the small triaxial region appears in the deformed area close
to $\chi=-\sqrt{7}/2$, we present in Fig.~\ref{fig-camino1b} a
path with fixed $\chi=-\sqrt{7}/2$ and $\xi=0.5$ and changing $k_3$.
Once more we plot in this figure the evolution of the
ground state energy and the equilibrium value of $\beta$ and
$\gamma$. Again low values of $k_3$ give rise to prolate shapes, while
the larger ones produce oblate forms, but a narrow region of triaxiality exists around 
$k_3\sim1.75-1.79$. The scale does not allow to discriminate, therefore we have 
enlarged it in the inset that shows two, very close, second order phase transitions. 
One from prolate to triaxial at around  $k_3=1.75$ and a second
one from triaxial to oblate at around $k_3=1.79$. As soon as we depart from
the $\chi=-\sqrt{7}/2$ surface these two second order surfaces approach
more and more up to a point in which both coincide and apparently transform into
a first order surface that directly separates prolate and oblate
shapes, without any triaxiality in between. It is very difficult
numerically to determine whether the triaxial region narrows 
indefinitely or rather ends up in a tricritical point.

\subsubsection{A path going from spherical through oblate and prolate shapes}

\begin{figure}
\begin{tabular}{cc}
\includegraphics[width=9cm , clip=]{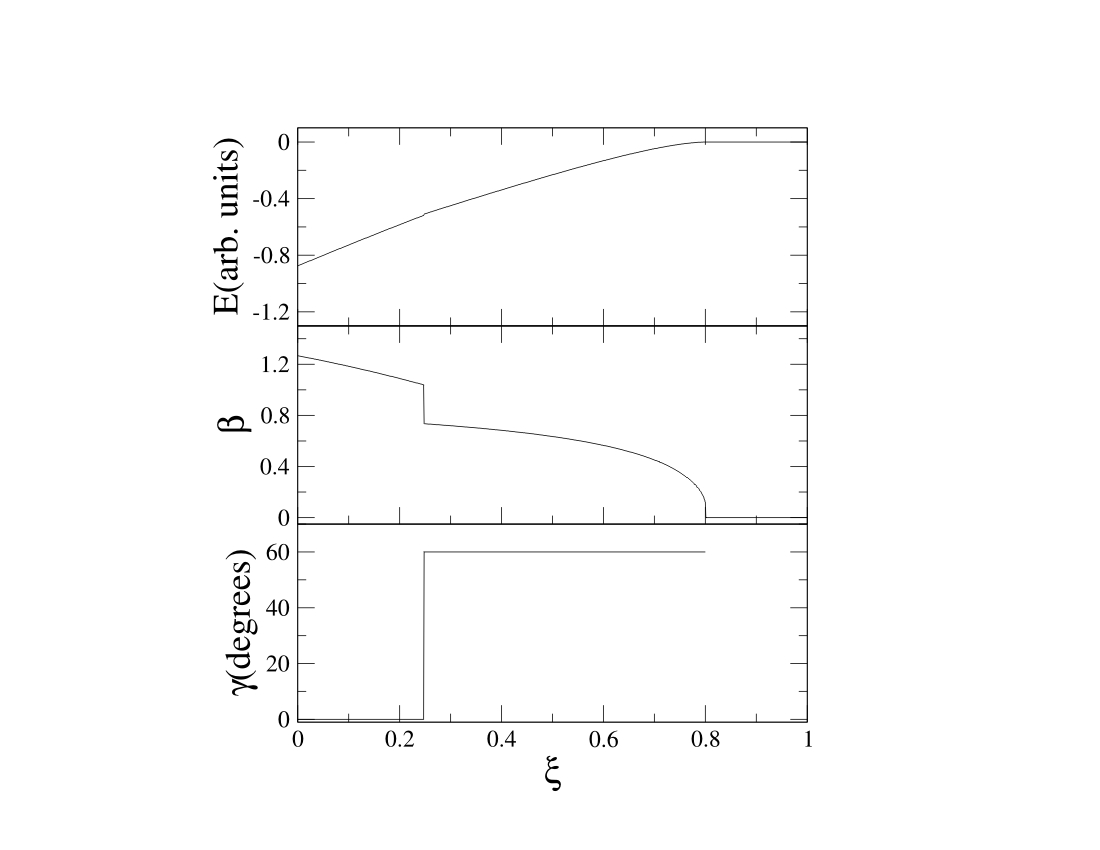}&
\includegraphics[width=4cm]{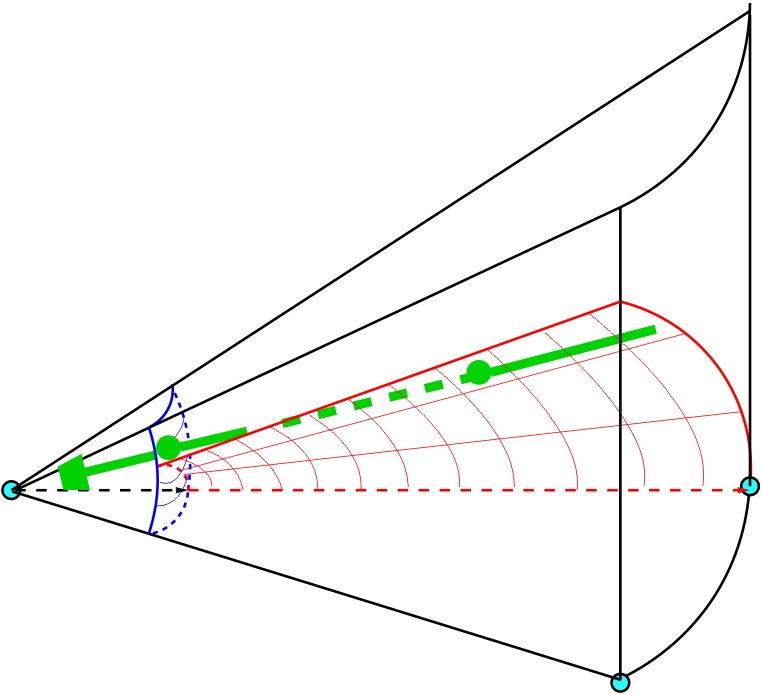}
\end{tabular}
\caption{(Color online) The same as in Fig.~\ref{fig-camino2} but for $k_3=1.5$ and $\chi=-2/\sqrt{5}$.} 
\label{fig-camino7}
\end{figure}

As a final calculation we study a trajectory that crosses the
spherical-deformed surface as well as the prolate-oblate one. 
The appropriate values of $k_3$ and $\chi$ are constrained by
equations (\ref{k_2o}) and (\ref{k_c}) and should verify   
\begin{equation}
-\frac{7 \sqrt{5} \chi }{2 \chi^2+7} >k_3>-\frac{\sqrt{5}}{2} \chi.  
\label{2ph}
\end{equation}
We choose
$k_3=1.5$ and $\chi=-2/\sqrt{5}$, which fulfil (\ref{2ph}), and vary the value of
$\xi$. This path is depicted in Fig.~\ref{fig-camino7} and two first
order phase transitions are clearly marked. From left to right, the
first one has a discontinuity in $\beta$ between two finite values
changing at the same time $\gamma$ from $0^\circ$ to $60^\circ$. In the second
one the system goes from the oblate phase to the spherical one,
changing $\beta$ from a finite to a zero value and $\gamma$ from
$60^\circ$ to undefined.
 
\section{Some applications }
\label{sec-appli}

\subsection{Spherical to axially deformed critical point: X(5)}
\begin{figure}[!ht]
\begin{center}
\begin{tabular}{cc}
\includegraphics[width=3cm , clip=]{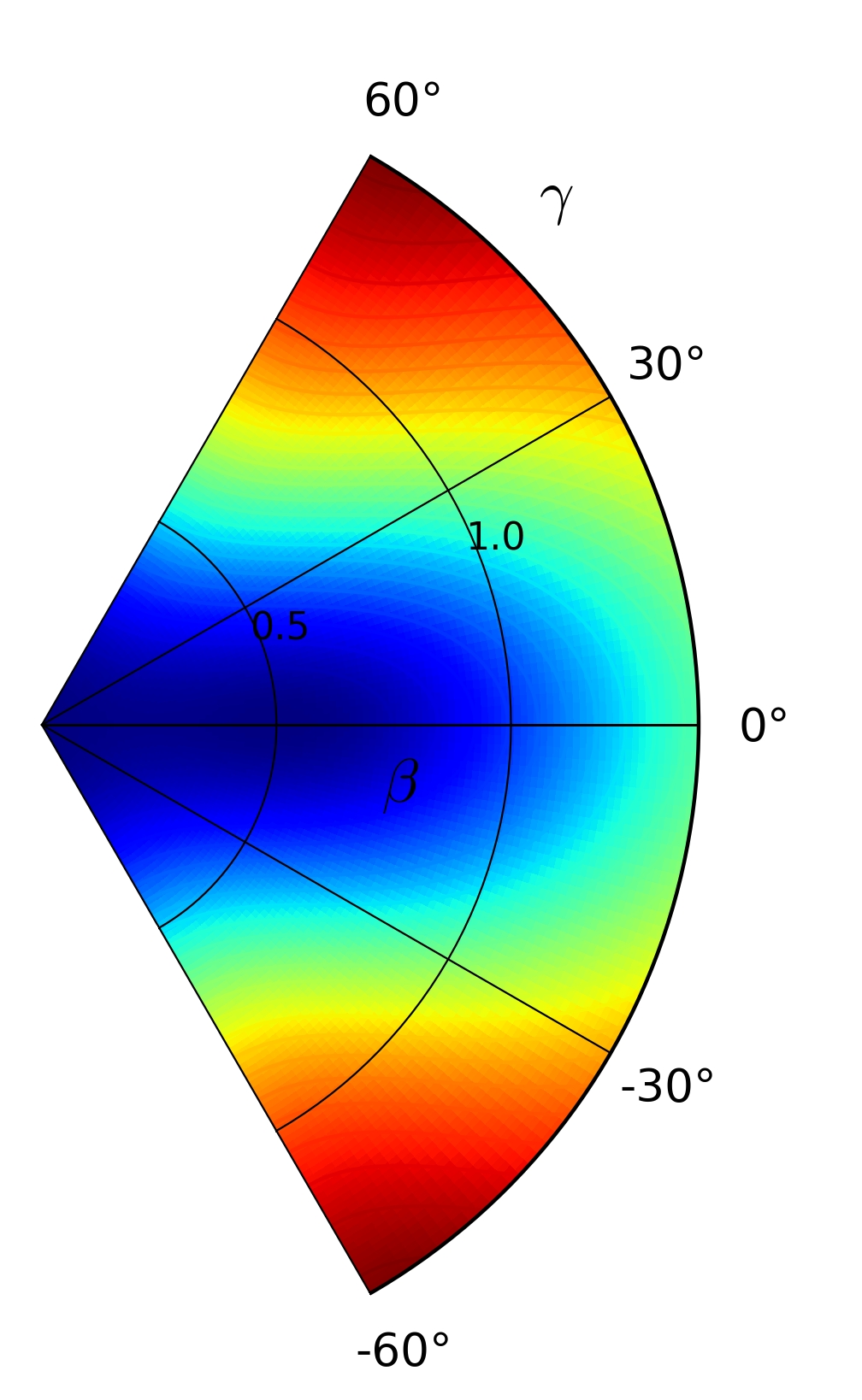}&
\includegraphics[width=4cm , clip=]{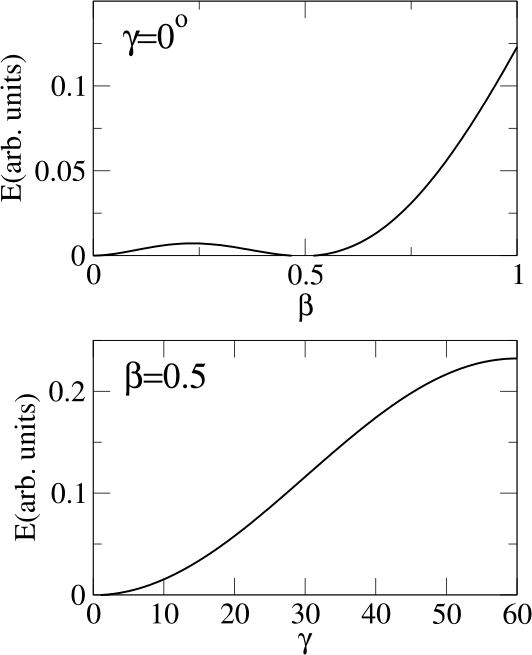}
\end{tabular}
\end{center}
\caption{(Color online) Potential energy surface (left) obtained from (\ref{h23}) with
  $\chi=0, \xi=0.9$ and $k_3=-9.5$. The valley along the $\gamma=0^\circ$
  direction for prolate deformations resembles the potential used in
  the X(5) solution of the Bohr Hamiltonian \cite{Iac2}.Right hand side: two sections
 of the energy surface around the equilibrium value of the deformation parameters 
have been plotted as a function of $\beta$ and $\gamma$, respectively.} 
\label{aX5}
\end{figure}
An interesting special case is to keep $\chi=0$,  in order to eliminate the
driving effect of the $\hat{Q}\cdot \hat{Q}$ term toward axial deformation. In the left part of 
Fig. \ref{aX5} a large value of $k_3$ generates a surface with an
almost flat valley along the $\beta$ direction, while the prolate
(oblate) confinement in $\gamma$ is due to the negative (positive)
sign in front of the $(\hat{Q}\times \hat{Q}\times \hat{Q})^{(0)}$ term. This case has a
potential energy surface that has a flat shape qualitatively similar
to the $V(\beta, \gamma)$ potential used in the X(5) critical point
symmetry \cite{Iac2}, with due differences: it is generated through a
set of parameters $\chi=0$, $\xi=0.9$, and $k_3=-9.5$ (a positive
value would generate an equivalent oblate shape), 
the potential does not tend to
infinity, but goes smoothly to an asymptotic value and the periodicity
in $\gamma$ is retained, 
at a variance with a harmonic oscillator. 
The bottom of the potential is not exactly flat, nor the behavior
approximates a $\beta^4$ dependence, but the fact that a potential
energy surface that mimics X(5) could be obtained with $\chi=0$ is
somewhat surprising, because it has been used to associate X(5) 
to a case that is intermediate between spherical and axially prolate 
shapes, while here we do not even need to set $\chi=-\sqrt{7}/2$.

\subsection{Application to a schematic Hamiltonian} 
\begin{figure*}[!t]
\begin{tabular}{cccc}
\includegraphics[clip=,width=0.24\textwidth, bb= 0 0 315 315]{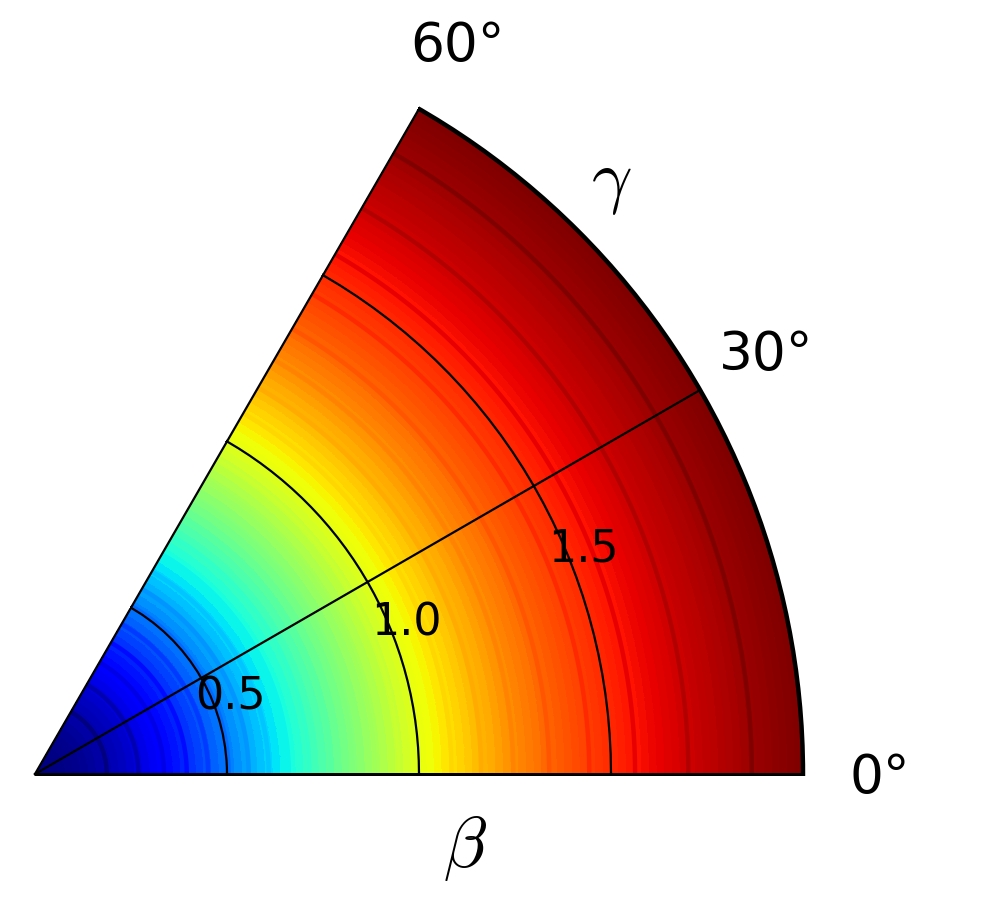} &
\includegraphics[clip=,width=0.24\textwidth, bb= 0 0 315 315]{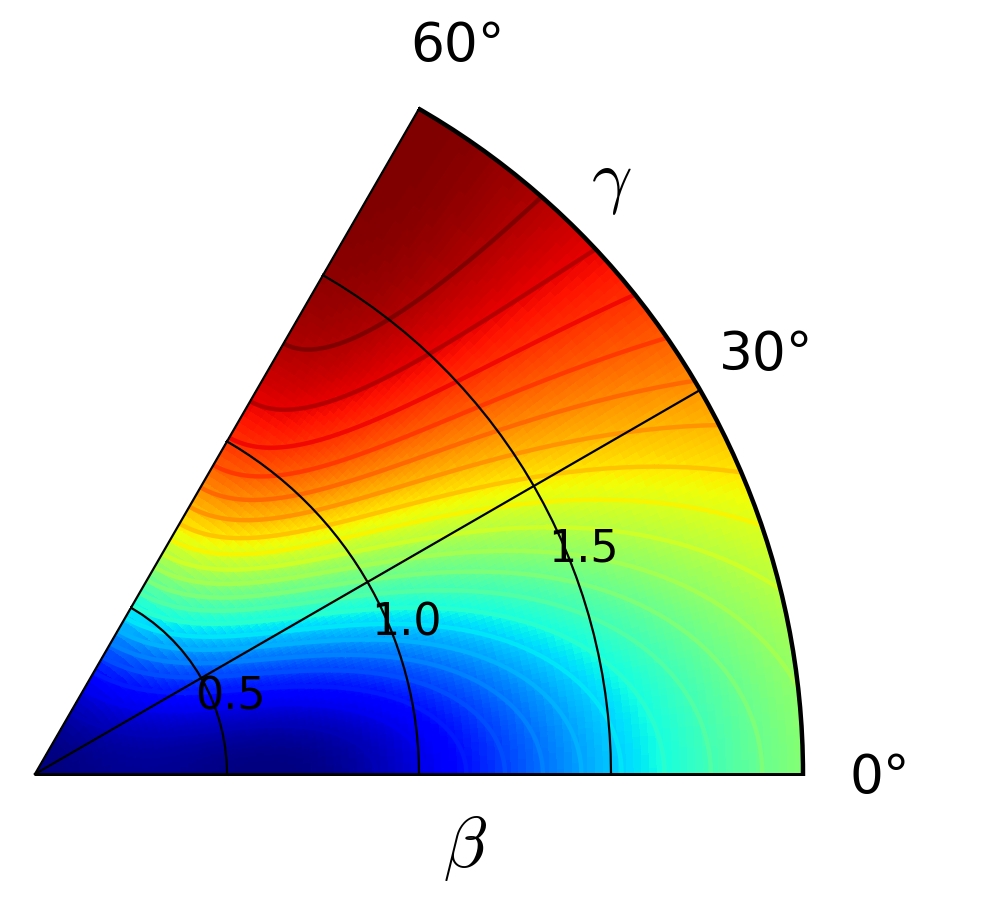}  &
\includegraphics[clip=,width=0.24\textwidth, bb= 0 0 315 315]{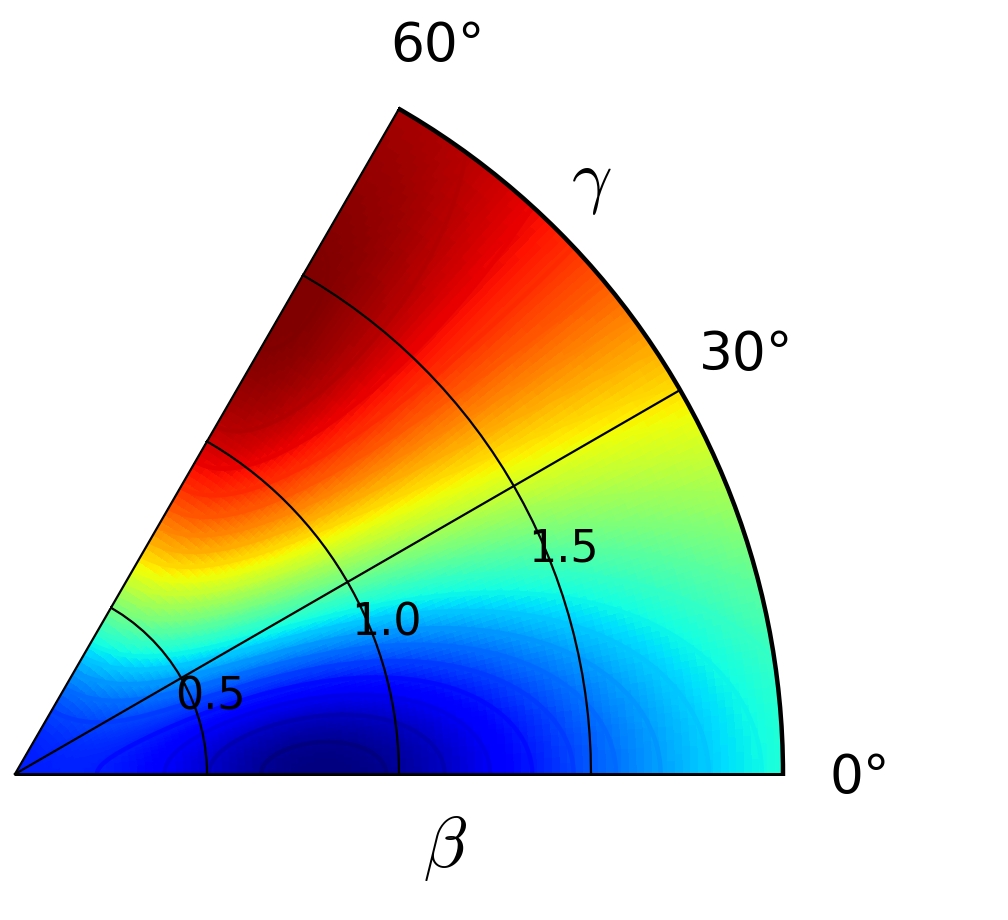}  &
\includegraphics[clip=,width=0.24\textwidth, bb= 0 0 315 315]{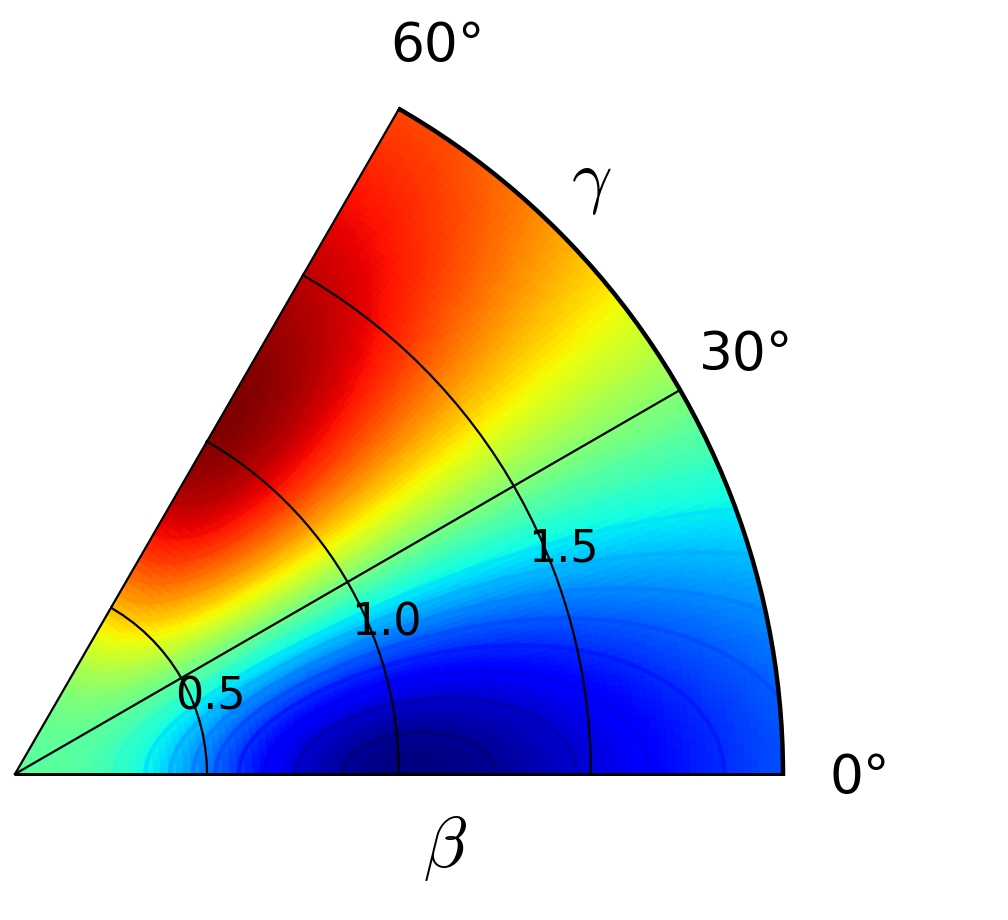} \\ 
k=0.0 & k$\sim$0.58776 &
k=1.0 & k=10.0 \\
\end{tabular}
\caption{(Color online) Potential energy surfaces for the Hamiltonian (\ref{h3}) with
  $N=10$ and different values of $k:0.0, k_{cr}, 1.0$ and 
$k=10.$. The minimum (blue) ranges from spherical to axially deformed prolate. 
Vertical scales change along the transition.} 
\label{string1}
\end{figure*}

\begin{figure}[!t]
\begin{center}
\includegraphics[width=15cm, clip=]{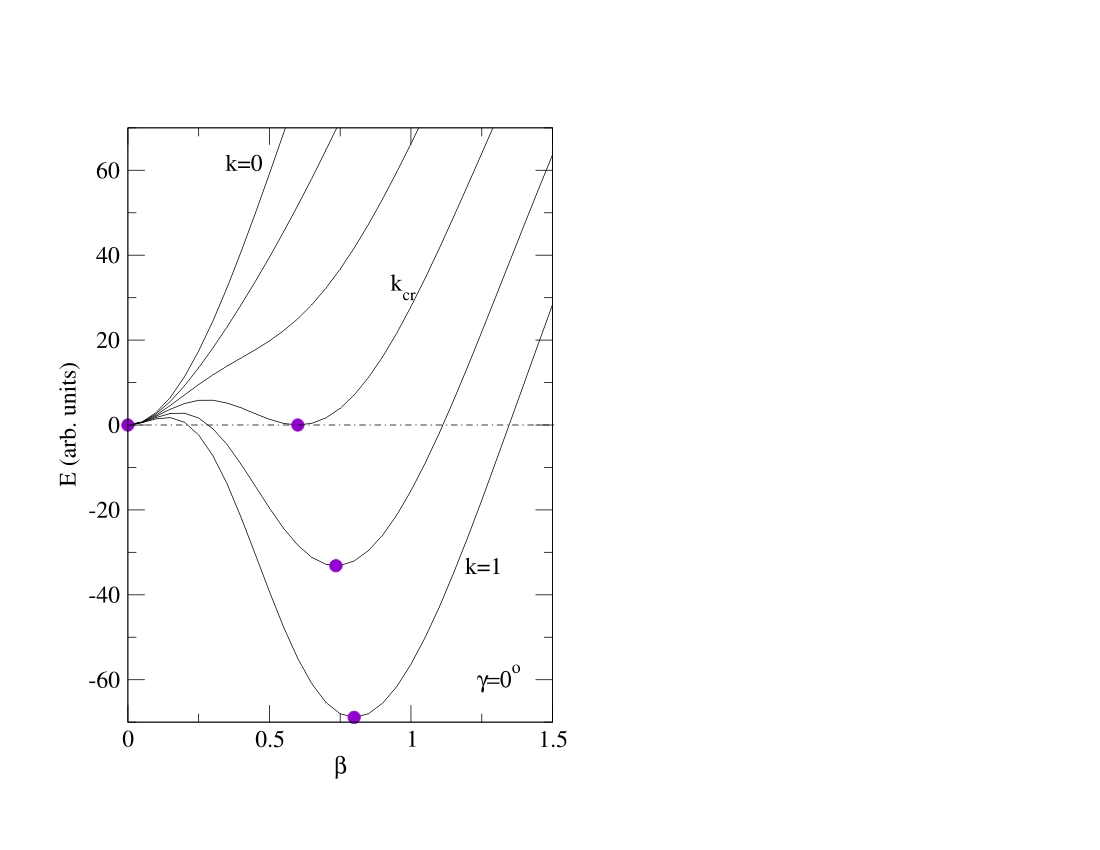}
\caption{(Color online) Cuts along $\gamma=0^\circ$ of the potential energy surfaces for the Hamiltonian
  (\ref{h3}) with $\chi=0$ and $N=10$ for several values of $k$ from
  $0.0$ to $1.0$. At the value  
$k_{cr}\sim 0.58776$ it shows the typical signs of a first order phase
transition. Purple dots indicate the position of minima.} 
\label{coll}
\end{center}
\end{figure}

As an application, we can now use the formula for the matrix element
of the $(\hat{Q}\times \hat{Q}\times \hat{Q})^{(0)}$ operator with $\chi~=~0$
within the boson coherent
state formalism to give further insight into the results obtained by
Rowe and Thiamova for the schematic Hamiltonian (\ref{h3}) \cite{RoTi,Thi} . The
parameter $k$ is allowed to vary from zero to large positive
values. In Fig. \ref{string1} we have plotted several contour maps
corresponding to $k=0.0,~ 0.58776,~ 1.0$, and $10.0$. Notice that the
$\Lambda$ term alone ($k=0$) just gives a spherical minimum, because
its matrix element in the coherent state approach is just proportional to
$\beta^2$.  
From the figure one can appreciate how, with growing value of $k$, the
minimum moves from the spherical configuration to the axially deformed
prolate one. This is due to the competition between the spherically
driving term $\Lambda$ and the cubic interaction term $(\hat{Q}\times
\hat{Q}\times \hat{Q})^{(0)}$ that has a stable axial minimum, despite having
$\chi=0$.  We have collected in Fig. \ref{coll} several cuts of the
potential energy surface along $\gamma=0^\circ$, for the values
$k=0.0,~ 0.2,~ 0.4,~ k_{cr},~ 0.8$, and $1.0$. Alongside the
ever-present minimum at $\beta=0$, this potential admits a second
deformed local minimum, that starts to appear after $k \sim 0.51$. The
critical point for the first order phase transition is found around
$k_{cr}\sim 0.58776$, where two minima are degenerate. After this point the
deformed minimum prevails. The values discussed above are valid with
the particular choice of $N=10$ and with the Hamiltonian as given
originally in (\ref{h3}). Although it is immaterial to the present
discussion, one should notice that, while the O(5) scalar term is
linear with the boson number, the $(\hat{Q}\times \hat{Q}\times \hat{Q})^{(0)}$ term is
cubic with $N$ and therefore one should better choose a Hamiltonian
that is properly normalized to get rid of the different scaling with
the boson number.
Notice that Ref.~\cite{piet} uses a slightly different Hamiltonian
having $\hat{L}\cdot \hat{L}$ as a spherical term: although it is most certainly
important to use the proper terms when calculating an IBM spectrum,
it is less critical here, since both these operators have matrix
elements proportional to $\beta^2$ (albeit not with the same
coefficients). In both cases \cite{piet,RoTi} triaxial minima are clearly not
possible. Moreover, the inclusion of the cubic term, even with
$\chi=0$, forbids the existence of a $\gamma-$independent solution.

\section{Conclusions}
\label{sec-conclu}

We have calculated the potential energy surface of the cubic $(\hat{Q}\times
\hat{Q}\times \hat{Q})^{(0)}$ term within the coherent state formalism with the
most general expression for the quadrupole operator. This has allowed
us to confirm the results of Ref. \cite{piet, RoTi} concerning the
fact that this term with $\chi=0$ can generate axially deformed
minima, but in addition it has been shown that the cubic term has already
a dependence on $\cos^2{3\gamma}$ (for $\chi\ne 0$) together with the dependence on $\cos{3\gamma}$
(obtainable also when $\chi=0$).  Therefore a Hamiltonian
containing this term may generate triaxiality without the
need to resort to the more complicated $(\hat{Q}\times \hat{Q}\times \hat{Q})^{(0)}
\cdot (\hat{Q}\times \hat{Q}\times \hat{Q})^{(0)}$ expression of Ref. \cite{Thi}.

After a discussion of its general features, we have used it to extend
the Consistent-Q Hamiltonian to the Cubic Consistent-Q Hamiltonian, of
which we have discussed the phase space, discovering that a triaxial
region can indeed be found, between the oblate and prolate phases.  
This region is extremely tiny (at least in the particular parameterization chosen, 
if compared to the other phases), and can be pinpointed numerically 
only very close to the limiting values of $\chi \sim \pm \sqrt{7}/2$. 
The inspection of the PES confirmed that the minimum is indeed triaxial 
inside this region. 
It is not clear, at the moment, if the prolate and oblate phases are always separated by a
region of triaxiality that progressively tails off as one goes to $\chi=0$ or 
if the triaxial region disappear at some point.
The numerical results suggest that this tiny triaxial region finishes
in a line where prolate, oblate and triaxial shapes coexist, {\it i.e.}~a
tricritical line.

\section*{Acknowledgments}
We acknowledge enlightening conversations on this topic with F. Iachello.
This work has been partially supported
by a Italian INFN-Spanish MCYT scientific agreement (ACI2009-1047 and AIC10-D-000590),
by the Spanish Ministerio de Educaci\'on y Ciencia and the European
regional development fund (FEDER) under projects number
FIS2008-04189 and FPA2007-63074, and by CPAN-Ingenio, by Junta de
Andaluc\'{\i}a under projects FQM160, FQM318, P05-FQM437 and
P07-FQM-02962.

\end{document}